\documentclass[structabstract]{aa}

\usepackage{graphicx}
\usepackage{amssymb,amsmath}
\usepackage{natbib}
\usepackage{txfonts}

\def\Msun{ M_\odot}

\def\Mp{M_{\rm p}}
\def\Rp{R_{\rm p}}
\def\Ms{M_{\ast}}

\def\Mearth{ M_\oplus}
\def\Rearth{ R_\oplus}
\def\Op{\Omega_{\rm p}}
\def\Os{\Omega_{\ast}}
\def\Tp{T_{\rm p}}
\def\Ts{T_\ast}
\def\G{\mathcal{G}}

\def\d{\mathrm{d}}
\def\i{\mathrm{i}}

\def\cm{\mathrm{cm}}
\def\m{\mathrm{m}}
\def\sec{\mathrm{s}}
\def\kg{\mathrm{kg}}
\def\gr{\mathrm{g}}

\begin{document}

\title{Tidal evolution of planets around brown dwarfs} 
   \subtitle{ }

   \author{ E. Bolmont \inst{1,2}
          \and S. N. Raymond \inst{1,2}  \and J. Leconte \inst{3}
                }

   \institute{Universit\'e de Bordeaux, Observatoire Aquitain des Sciences de l'Univers, 2 rue de l'Observatoire, BP 89, F-33271 Floirac Cedex, France
   			 \and CNRS, UMR 5804, Laboratoire d'Astrophysique de Bordeaux, 2 rue de l'Observatoire, BP 89, F-33271 Floirac Cedex, France 
%              \email{Ugo.Hincelin@obs.u-bordeaux1.fr}
         \and \'Ecole Normale Sup\'erieure de Lyon, CRAL (CNRS), Universit\'e de Lyon, 46 all\'ee d'Italie, 69007 Lyon, France}
 %            \email{c.ptolemy@hipparch.uheaven.space}
%             \thanks{The university of heaven temporarily does not
  %                   accept e-mails}

   \date{Received xxx ; accepted xxx}

% \abstract{}{}{}{}{}
% 5 {} token are mandatory

  \abstract
   {The tidal evolution of planets orbiting brown dwarfs (BDs) presents an interesting case study because BDs' terrestrial planet forming region is located extremely close-in. In fact, the habitable zones of BDs range from roughly $0.001$ to $0.03$~AU and for the lowest-mass BDs are located interior to the Roche limit.}
   {In contrast with stars, BDs spin up as they age. Thus, the corotation distance moves inward. This has important implications for the tidal evolution of planets around BDs.}
   {We used a standard equilibrium tidal model to compute the orbital evolution of a large ensemble of planet-BD systems. We tested the effect of numerous parameters such as the initial semi-major axis and eccentricity, the rotation period of the BD, the masses of both the BD and planet, and the tidal dissipation factors.}
   {We find that all planets that form at or beyond the corotation distance and with initial eccentricities smaller than $\sim 0.1$ are repelled from the BD.  Some planets initially interior to corotation can survive if their inward tidal evolution is slower than the BD's spin evolution, but most initially close-in planets fall onto the BD. }
   {We find that the most important parameter for the tidal evolution is the initial orbital distance with respect to the corotation distance. Some planets can survive in the habitable zone for Gyr timescales, although in many cases the habitable zone moves inward past the planet's orbit in just tens to hundreds of Myr. Surviving planets can have orbital periods of less than $10$~days (as small as $10$~hrs), so they could be observable by transit.}
		
   \keywords{%{\bf SUPERZOYOTTE
   	Brown dwarfs --
                Stars: rotation --
                Planets and satellites: dynamical evolution and stability --
                Planet-star interactions --
                Astrobiology}
   
\maketitle
%
%________________________________________________________________

\section{Introduction}

Tidal effects around brown dwarfs (BDs) have received little attention but they present an interesting case study because the planets are thought to form very close to their star in a region where tides are very strong and lead to significant orbital changes and extreme internal planetary heating \citep[see][]{PapaloizouTerquem2010, Barnes2010}. Moreover, as the luminosity of brown dwarfs decreases dramatically on Gyr timescales \citep[see][]{Tout1999}, the habitable zone moves inward in time and meets the formation region of planets, which naturally raises the question of habitability of these planets.  

The parameter that governs the direction of tidal evolution for a planet orbiting a star or BD (or a satellite orbiting a planet) is the initial semi-major axis with respect to the corotation radius, the orbital radius where the orbital period matches the central body's spin period. For a planet interior to the corotation radius, the planet's mean motion is slower than the primary's rotation so the tidal bulge raised by the planet on the primary lags behind the position of the planet.  The planet feels a drag force that slows it down and causes its orbital radius to shrink, in some cases leading to an eventual merger with the primary. However, for a planet exterior to the corotation radius,  tidal forces push the planet outward because the tidal bulge on the star is in advance with respect to the position of the planet.

Main sequence stars spin down in time due to angular momentum loss driven by stellar winds \citep{Skumanich1972, Matt2010}.  In contrast, BDs spin up with time \citep[see][]{Herbst2007, ScholzEisloffel2005}.  This key difference, which is due to the low efficiency of the magnetic braking in the substellar domain (see Section 2.3.3), is of great importance for the tidal evolution of planets orbiting BDs because the primary's spin rate sets the position of the corotation radius. For stars, the corotation distance expands as the star spins down, so close-in planets spiral inward and, in time, more distant planets also start to spiral inward.  However, the corotation distance around BDs shrinks in time such that some planets may spiral inward at early times but later, when the corotation radius moves interior to their orbits, change direction and spiral outward, away from the BD.

The habitability of planets around BDs also differs from planets around main sequence stars.  Given the inability of BDs to fuse H into He, BDs simply cool in time and their habitable zones move inward.  For a planet on a fixed orbit, the habitable zone will thus sweep past it in a finite time~\citep{Andreeshchev2004}.  For a planet being pushed outward by tides, the ``habitable interval'' will be even shorter. The time a planet will spend in the habitable zone depends on many parameters, the main one being the mass of the star.

Furthermore, BDsÕ small sizes and masses also make them excellent targets for radial velocity and transit searches \citep[see][]{Blake2008}, although significant observational constraints exist (e.g., the difficulty of measuring precise radial velocities in the infrared Ð \citep[see][]{Blake2010}). BDs are thought to be more numerous than low mass stars, and recently five new BDs have been discovered at $\sim 10$~pc \citep{Burgasser2011} and two at $\sim 5$~pc \citep{Scholz2011} with WISE increasing the number of targets for observation of planetary transit. 

There are several reasons to think that planets should form around BDs.  The disk fraction of young BDs is comparable to that for T Tauri stars \citep[see][]{Jayawardhana2003, Luhman2005}, and grain growth and settling have been observed in disks around BDs \citep{Apai2005}.  Protoplanetary disk masses are thought to correlate roughly linearly with the stellar mass \citep{Scholz2006, Andrews2007}, such that BD disks might be systematically lower-mass than those around main sequence stars. However, recent sub-mm observations suggest that there exists a correlation between protoplanetary disk mass and radial extent such that lower-mass disks are more compact \citep{Andrews2010}, and disks around low-mass stars may actually last longer than around more massive stars \citep{Pascucci2008}.  In addition, Kepler observations have found that low-mass stars actually harbor close-in super Earth-sized planets at a higher frequency than Sun-like stars \citep{Howard2011}.  Thus, we expect that there probably exists a sizable niche in the parameter space of BD and disk properties (BD mass, disk mass, surface density distribution, and lifetime) that allows for the formation of sizeable ($\sim0.1$ Earth masses or larger) terrestrial planets near the habitable zone of BDs \citep[see][]{Raymond2007,PayneLodato2007}.

\medskip

Here we use a standard equilibrium model to study the tidal evolution of planets orbiting BDs.  Our paper is structured as follows. The tidal and BD evolutionary models are discussed in Section \ref{model1}.  The results of the tidal evolution of planets around BDs and the influence of several parameters are presented in Section \ref{Results}. Finally, in Section \ref{Discussion} we discuss tidal evolution with regards to the habitable zone around BDs and the importance of tidal evolution for observations.

%%%%%%%%%%%%%%%%%%%%%%%%%%%%%%%%%%%%%%%%%%%%%%%%%%%%%%%%%%%%%%%
%%%%%%%%%%%%%%%%%%%%%%%%%%%%%%%%%%%%%%%%%%%%%%%%%%%%%%%%%%%%%%%

\section{Model description}
\label{model1}

We have developed a model to study the orbital evolution of planets around BDs by solving the tidal equations for arbitrary eccentricity and also taking into account the observed spin evolution of BDs.

%%%%%%%%%%%%%%%%%%%%%%%%%%%%%%%%%%%%%%%%%%%%%%%%%%%%%%%%%%%%%%%
\subsection{Tidal model}

The tidal model that we used is a re-derivation of the equilibrium tide model of \citet{Hut1981} as in \citet{EKH1998}. We consider both the tide raised by the BD on the planet and by the planet on the BD. We use the constant time lag model \citep{Leconte2010} and use the internal dissipation constant $\sigma$ which was calibrated for giant exoplanets and their host stars by \citet{Hansen2010}. 

\medskip

The secular tidal evolution of the semi-major axis $a$ is given by \citep{Hansen2010}: 
\begin{equation}\label{Hansena}
\begin{split}
\frac{1}{a}\frac{\d a}{\d t} &= -\frac{1}{\Tp}\Big[Na1(e)-\frac{\Op}{n}Na2(e)\Big] \\
& \quad - \frac{1}{\Ts}\Big[Na1(e)-\frac{\Os}{n}Na2(e)\Big],
\end{split}
\end{equation}

where the dissipation timescale $\Tp$ is defined as
\begin{equation}
\label{Tp}
\Tp = \frac{1}{9}\frac{\Mp}{\Ms(\Mp+\Ms)}\frac{a^8}{\Rp^{10}}\frac{1}{\sigma_{p}}
\end{equation}
and depends on the mass of the planet $\Mp$, its dissipation $\sigma_{p}$ and of the mass of the BD $\Ms$. $\Op$ is the planet rotation frequency, and $n$ is the mean orbital angular frequency. The stellar parameters are obtained by switching the $p$ and $\ast$ indices. $Na1(e)$ and $Na2(e)$ are eccentricity-dependent factors, which are valid even for very high eccentricity \citep{Leconte2010} : 
\begin{align*}
Na1(e) &= \frac{1+31/2e^2+255/8e^4+185/16e^6+85/64e^8}{(1-e^{2})^{15/2}},\\
Na2(e) &= \frac{1+15/2e^2+45/8e^4+5/16e^6}{(1-e^{2})^{6}}.
\end{align*}

Similarly, the evolution of the eccentricity of the planet is given by : 

\begin{equation}\label{Hansene}
\begin{split}
\frac{1}{e}\frac{\mathrm{d}e}{\mathrm{d}t} &= -\frac{9}{2\Tp}\Big[Ne1(e)-\frac{11}{18}\frac{\Op}{n}Ne2(e)\Big]\\ 
& \quad - \frac{9}{2\Ts}\Big[Ne1(e)-\frac{11}{18}\frac{\Os}{n}Ne2(e)\Big],
\end{split}
\end{equation}

with 
\begin{align*}
Ne1(e) &= \frac{1+15/4e^2+15/8e^4+5/64e^6}{(1-e^{2})^{13/2}},\\
Ne2(e) &= \frac{1+3/2e^2+1/8e^4}{(1-e^{2})^{5}}.
\end{align*}
   
Finally, the rotational state of each object is given by : 
\begin{align}
\label{Hansenop}
\frac{1}{\Op}\frac{\mathrm{d}\Op}{\mathrm{d}t} &= -\frac{\gamma_{p}}{2\Tp}\Big[No1(e)-\frac{\Op}{n}No2(e)\Big],
\end{align}
and
\begin{equation}\label{Hansenos}
\begin{split}
\frac{1}{\Os}\frac{\mathrm{d}\Os}{\mathrm{d}t} &= -\frac{\gamma_{\ast}}{2\Ts}\Big[No1(e)-\frac{\Os}{n}No2(e)\Big] \\
& \quad - \frac{2}{R_{\ast}}\frac{\mathrm{d}R_{\ast}}{\mathrm{d}t}-\frac{1}{rg2_{\ast}}\frac{\mathrm{d}rg2_{\ast}}{\mathrm{d}t},
\end{split}
\end{equation}

Here, $\gamma_{i}=\frac{h}{I_i\Omega_i}$ is the ratio of orbital angular momentum $h$ to spin angular momentum and $rg2_{\ast}$ is the square of the parameter $rg_{\ast}$ (which is the radius of gyration of \citet{Hut1981}) which is defined as : $I=\Ms(rg_{\ast}R_{\ast})^2$, where $I$ is the moment of inertia of the BD, and
\begin{align*}
No1(e) &= \frac{1+15/2e^2+45/8e^4+5/16e^6}{(1-e^{2})^{13/2}},\\
No2(e) &= \frac{1+3e^2+3/8e^4}{(1-e^{2})^{5}}.
\end{align*}

The last two terms of equation (\ref{Hansenos}) translate the conservation of the moment of inertia. As will be discussed in the next subsection, in the case of BDs the radius decreases with time.

\medskip

The integration of these equations was performed using a fourth order Runge-Kutta integrator with an adaptive timestep routine \citep{Press1992}. The precision of the calculations was chosen such that the final semi-major axis of each integrated system was robust to numerical error at a level of at most one part in $10^3$.

%%%%%%%%%%%%%%%%%%%%%%%%%%%%%%%%%%%%%%%%%%%%%%%%%%%%%%%%%%%%%%%

\subsection{Planets model}

Given the relatively small mass of protoplanetary disks around BDs, we assume that most planets will be terrestrial planets rather than gas giants. Although we do not exclude the possible existence of gas giants around BDs (see Section \ref{infplanetmass}), we focus on terrestrial planets whose internal dissipation factor is the same as Earth's but scaled with the radius of the planet. 
 
 \citet{DeSurgyLaskar1997} inferred the quantity $k_{2,\oplus}\Delta t_{\oplus} = 213$~s from the DE245 data for Earth. $k_{2,p}$ is the Earth's potential Love number of degree $2$, which is a parameter depending on the moment of inertia of the body. It tells us how the body responds to compression ($k_2 = 3/2$ means that the body is an incompressible ideal fluid planet). $\Delta \Tp$ is the constant time-lag. Hansen's $\sigma_{p}$ and the quantity $k_{2,p}\Delta \Tp$ are related through : 
\begin{equation}
\label{kdtsigma}
k_{2,p}\Delta \Tp = \frac{3}{2}\frac{\Rp^{5}\sigma_{p}}{\G},
\end{equation}

where $\Rp$ is the planetary radius and $\G$ is the gravitational constant.

The planet's compositions are assumed to be an Earth-like mixture of rock and iron, following the mass-radius relation of \citet{Fortney2007}: 

\begin{equation}
\begin{split}
\Rp/\Rearth &= (0.0592~rmf + 0975)~\Big(log\frac{\Mp}{\Mearth}\Big)^2\\
&  \quad+ (0.2337~rmf + 0.4938)~log\frac{\Mp}{\Mearth} \\
&  \quad+ (0.3102~rmf + 0.7932),
\end{split}
\end{equation}

where $rmf$ is the rock mass fraction. For Earth, $rmf \sim 0.7$.

Concerning the evolution of the planet it is interesting to note that Equation (\ref{Hansenop}) has a much shorter timescale than the other differential equations. Very quickly in its evolution the planet reaches pseudo-synchronization, which means that its rotation tends to be synchronized with the orbital angular velocity at periastron \citep{Hut1981}. Indeed, the tidal interaction is strongest near periastron where the planet is closest to the BD. Having verified this behavior in preliminary simulations, the rotation period is fixed to the pseudo-synchronization value at every calculation timestep.

In this study, we chose not to treat the evolution of the obliquity of the bodies for simplicity. \citet{HellerLeconteBarnes2011} showed that the timescale of evolution of obliquity  -- which they call ``tilt erosion`` time -- is shorter for low mass stars and for close-in planets. BDs have low masses and the planets we consider in this study are very close-in, so we can assume that the obliquity is negligible.  

%%%%%%%%%%%%%%%%%%%%%%%%%%%%%%%%%%%%%%%%%%%%%%%%%%%%%%%%%%%%%%%%%%%%%%%%%%%%%%%%%%%%%%%%%%%%%%%%%%%%%%%%%%%%%%%%%%%%%%%%%%%%%%

\subsection{Brown Dwarf Evolution}

BDs are self-gravitating objects that formed through gravitational collapse of a dense molecular cloud but that are not hot enough to ignite proton fusion (REF BD review). As the hydrogen burning temperature is around $3\times10^6\,$K, only objects more massive than $M_\mathrm{HBMM}\approx 75~M_{JUP}$ are able to start the PPI nuclear reaction chain, providing us with an upper bound for the BD mass domain \citep{ChabrierBaraffe1997, ChabrierBaraffe2000}. If the lower bound is more uncertain, both observational \citep{Caballero2007} and analytical \citep{PadoanNordlund2004, HennebelleChabrier2008} arguments suggest that the same formation process responsible for star formation can continuously produce objects down to a few Jupiter masses.
\textit{Mini brown dwarfs} and \textit{giant planets} (formed in a protoplanetary disk) thus overlap in mass, stressing the need for identification criteria enabling the distinction between these two types of astrophysical bodies \citep{Leconte2009, Spiegel2011, Leconte2011a}.

%%%%%%%%%%%%%%%%%%%%%%%%%%%%%%%

\subsubsection{Internal structure and luminosity}

To calculate the influence of the dissipation into the BD on the tidal evolution of the system, one needs to know the internal structure of the object and how it responds to a tidal perturbation, namely, the BD radius, moment of inertia (the gyration radius), tidal response (the love number, $k_2$) and tidal dissipation ($\sigma$).
The BD radius, moment of inertia and tidal response are provided by the grids of BDs and giant planets evolutionary models of \citet{Leconte2011}\footnote{Electronic versions of the model grids are available at \textit{http://perso.ens-lyon.fr/jeremy.leconte/JLSite/JLsite /Exoplanets\_Simulations.html}}. While the main physics inputs (equations of state, internal composition, atmosphere models, boundary conditions) used in these calculations have been described in
detail in previous papers devoted to the evolution of substellar objects \citep{ChabrierBaraffe1997, BaraffeChabrierBarman2003, Leconte2011} details the calculation of the dynamical properties of gaseous objects in hydrostatic equilibrium (inertia and love number).

Because most of the BD's interior remains fully convective for most of its lifetime, there is no drastic change in the dimensionless moment of inertia of the object with time, as is the case in young solar type stars when the radiative core develops. The main variation of the moment of inertia of the BD is due to its monotonous contraction caused by the radiative energy losses. However, if the object is more massive than $\sim0.0125~\Msun$ the temperature reached in the interior due to the early contraction is sufficient to ignite Deuterium burning (this limit is $\sim0.06~M_\odot$ for Lithium burning). As this additional energy source is temporarily able to nearly balance the radiative losses, the contraction is slowed down and the radius, the luminosity, and the moment of inertia reach a plateau. Because the primordial abundance of these compounds is small, this phase lasts only up to a few million years for massive brown dwarfs, and $100$~Myr near the deuterium burning limit.

%In our calculation we took into account the contraction of the BD, as seen in Equation \ref{Hansenos}. For convenience,  
%this equation can be partially integrated and computed as follows : 
%\begin{equation}\label{Hansenosbetter}
%%\begin{split}
%\Omega_{BD}(t) = \Omega_{BD}(t_0) \left[\frac{rg2_{BD}(t_0)}{rg2_{BD}(t)}\left(\frac{R_{BD}(t_0)}{R_{BD}(t)}\right)^2 \times \mathrm{exp}\left(\int_{t_0}^{t}f_{tides} \d t\right)\right],
%%\end{split}
%\end{equation}
%where $t_0$ is the initial time and $f_{tides}$ is given by : 
%\begin{equation}\label{Hansenosftides}
%\begin{split}
%f_{tides} & = \frac{1}{\Omega_{BD}}\frac{\d\Omega_{BD}}{\d t}\Big|_{R_{BD}=cst, rg2_{BD}=cst} \\
%&= -\frac{\gamma_{\ast}}{2T_{BD}}\Big[No1(e)-\frac{\Omega_{BD}}{n}No2(e)\Big].
%\end{split}
%\end{equation}
%
%For terrestrial planets, $f_{tides}$ remains small throughout the evolution, and the BD rotation period is mainly determined by the conservation of angular momentum and so by the initial rotation period.

To quantify the extent of the habitable zone we use a simple criterion based on the flux at the planet's substellar point. At a given semi major axis and eccentricity, this mean bolometric flux is 
\begin{align}
F_\mathrm{inc}=\left(\frac{R_{BD}}{a}\right)^2\,\frac{\sigma_\mathrm{SB} T_\mathrm{eff}^4}{\sqrt{1-e^2}},
\end{align}
where $\sigma_\mathrm{SB}$ is the Stefan Boltzmann constant, and $R_{BD}$, and $T_\mathrm{eff}$ are the radius and effective temperature of the BD.
 
%%%%%%%%%%%%%%%%%%%%%%%%%%%%%%%

\subsubsection{BD dissipation}

\citet{Hansen2010} makes the assumption that all substellar objects have the same internal dissipation because BDs and gas giants have similar internal structure and dissipative response. Note, however, that if the presence of a central dense (not necessarily solid) core plays a major role in the process responsible for the tidal dissipation in giant planets \citep{GoodmanLackner2009}, this assumption may not hold anymore, as BDs, which form like stars are not expected to possess a metal enriched core. In our notation, the internal dissipation factor $\sigma_{BD}$ is thus the same as $\sigma_{p}$ from \citet{Hansen2010}. 

The dissipation factor is poorly constrained but \citet{Hansen2010} gives estimates for stars : $\sigma_{\ast} = 6.4\times10^{-59}~\gr^{-1}\cm^{-2}\sec^{-1} \overline\sigma_{\ast}$ where $\overline\sigma_{\ast} = 7.8\times10^{-8}$ and for gas giants : $\sigma_{p} = 5.9\times10^{-54} ~\gr^{-1}\cm^{-2}\sec^{-1}\overline\sigma_p$ where $\overline\sigma_p = 3.4\times10^{-7}$. Thus for stars,  the dissipation factor is $\sigma_{\ast} = 4.992 \times 10^{-76}~\gr^{-1}\cm^{-2}\sec^{-1}$ and for gas giants or BDs it is $\sigma_{p} = 2.006 \times 10^{-60}~\gr^{-1}\cm^{-2}\sec^{-1}$.

So the value of the dissipation factor for the BDs thereafter referred as $\sigma_{BD}$ is : 

\begin{equation}\label{sigmas}
\sigma_{BD} = 2.006 \times 10^{-53}~\kg^{-1}\m^{-2}\sec^{-1}.
 \end{equation}

%%%%%%%%%%%%%%%%%%%%%%%%%%%%%%%

\subsubsection{Rotational angular velocity}

Brown dwarfs are very rapid rotators, which is surprising at first glance. The rapid rotation coupled to the convective interior creates a strong magnetic field which should entail an important magnetic braking through ejection of a magnetized wind. However, as shown by \citet{Mohanty2002}, in BDs' cool atmospheres the degree of ionization is very low. A combination of very low ionization fraction and high densities results in very large resistivities and thus efficient magnetic field diffusion. The magnetic braking is thus very inefficient in the substellar domain and cannot counteract the early spin up caused by the contraction. Thus, throughout this study, magnetic braking is neglected.

Information about the rotation periods of BDs comes from both statistical studies of BDs in open clusters \citep[see][and references therein]{Herbst2007}, for which ages are known, and specific studies of particular BDs, which can be studied in detail but for which age determinations are often challenging. 

BDs have observed orbital periods as short as a few hours \citep{Herbst2007}.  For example, the nearby BD DENIS-P J0041353-562112 has a rotation period of $2.8$~hr, an age between $10$ and $15$~Myr and a mass below $0.062~\Msun$ \citep{Koen2010}. The rotation period depends on the age of the BD and on its mass \citep{Herbst2007}. For a given age, low-mass BDs rotate faster than high-mass BDs, and all BDs spin slower at younger ages.  This behavior can be explained by the simple conservation of moment of inertia during BD contraction. 

\begin{figure}[h!]
	\begin{center}
	\includegraphics[width=8.5cm]{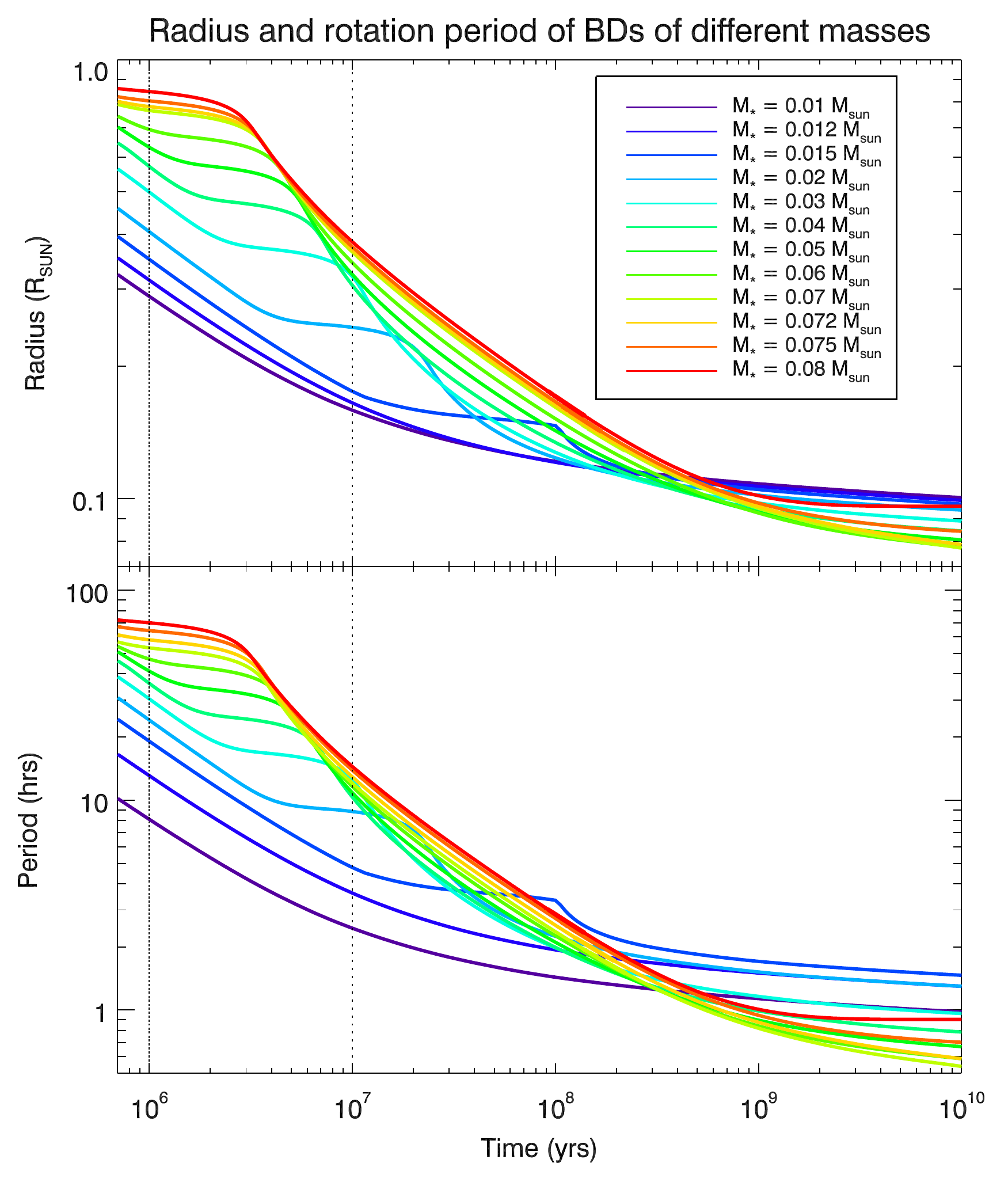}
	\caption{Spin evolution of BDs assumed in our simulations, calibrated to match observations \citep{Herbst2007}. The curves represent the rotation period of BDs of different masses ($0.01$, $0.012$, $0.015$, $0.02$, $0.03$, $0.04$, $0.05$, $0.06$, $0.07$, $0.072$, $0.075$, $0.08~\Msun$) computed from the radius evolution and the conservation of angular momentum.  The dotted lines represent the initial time for the two sets of simulations, thus also giving the initial rotation periods for each BD.}
	\label{rotperiodBD_article}
	\end{center}
	\end{figure}

The assumed evolution of the spin period of BDs adopted in our simulations is shown in Fig. \ref{rotperiodBD_article}.  After 1 million years, BD rotation periods are distributed between $8$~hr and $70$~hr  with longer initial periods for more massive BDs (the $70$~hr period corresponds to the highest mass BD of $0.08~\Msun$). The fact that higher-mass BDs have slower initial spin rates is an empirically-calibrated relation \citep{Herbst2007}, although the data are relatively scarce. We chose these initial values to match the order of magnitudes seen in \citet{Herbst2007} and we ensured that the final rotation periods were not too close to the break up period of the BD (i.e., the Keplerian period of a massless particle at the surface of the BD). Considering the uncertainty on the dissipation factor and the strong gravity at the surface, we ignore the increase of the mean radius of the BD caused by the fast rotation (see Fig. 1 of \citet{Leconte2011}).%We also assume that BDs remain spherical in shape throughout their evolution despite their fast rotation.

	We note that these initial conditions remain somewhat uncertain because the BD radii are uncertain at young ages. 
	
	In our calculation we took into account the contraction of the BD, as seen in Equation \ref{Hansenos}. For convenience,  
this equation can be partially integrated and computed as follows : 
\begin{equation}\label{Hansenosbetter}
%\begin{split}
\Omega_{BD}(t) = \Omega_{BD}(t_0) \left[\frac{rg2_{BD}(t_0)}{rg2_{BD}(t)}\left(\frac{R_{BD}(t_0)}{R_{BD}(t)}\right)^2 \times \mathrm{exp}\left(\int_{t_0}^{t}f_{tides} \d t\right)\right],
%\end{split}
\end{equation}
where $t_0$ is the initial time and $f_{tides}$ is given by : 
\begin{equation}\label{Hansenosftides}
\begin{split}
f_{tides} & = \frac{1}{\Omega_{BD}}\frac{\d\Omega_{BD}}{\d t}\Big|_{R_{BD}=cst, rg2_{BD}=cst} \\
&= -\frac{\gamma_{\ast}}{2T_{BD}}\Big[No1(e)-\frac{\Omega_{BD}}{n}No2(e)\Big].
\end{split}
\end{equation}

For terrestrial planets, $f_{tides}$ remains small throughout the evolution, and the BD rotation period is mainly determined by the conservation of angular momentum and so by the initial rotation period.

%%%%%%%%%%%%%%%%%%%%%%%%%%%%%%%%%%%%%%%%%%%%%%%%%%%%%%%%%%%%%%%
%%%%%%%%%%%%%%%%%%%%%%%%%%%%%%%%%%%%%%%%%%%%%%%%%%%%%%%%%%%%%%%

\section{Results}
\label{Results}
%%%%%%%%%%%%%%%%%%%%%%%%%%%%%%%%%%%%%%%%%%%%%%%%%%%%%%%%%%%%%%%

\subsection{An example}

We now present an example of the tidal evolution of planets around a BD. We consider a BD of mass $0.04~\Msun$, and a planet of $1~\Mearth$ placed at initial orbital radii between $7\times10^{-4}$~AU and $2.4\times10^{-3}$~AU. The BD dissipation is the same as used by \citet{Hansen2010} as given in equation (\ref{sigmas}). The planet has the same $k_{2,p}\Delta t_p$ ans $\sigma_p$ as Earth. The initial BD rotation period is $36$~hr and the initial orbital eccentricity is $0.01$.  The initial time is $10^6$~yrs, meaning that the BD's structure at the start of the integration was taken from a $1$~Myr snapshot in the BD models described above.  This means that we consider that the planet was  fully-formed at the time of the dissipation of the disk, which is consistent with simple estimates of the accretion time around very low-mass stars \citep{Raymond2007}.  Of course, given that gaseous protoplanetary disks have lifetimes of a few Myr \citep[see][]{Haisch2001, Hillenbrand2008} and that disks around low-mass stars tend to be longer-lived than around Sun-like stars \citep{Pascucci2009}, values for ``time zero'' as long as $10$~Myr are probably realistic (see Section \ref{inftime}).

\smallskip
 
\begin{figure}[h!]
	\begin{center}
	\includegraphics[width=8cm]{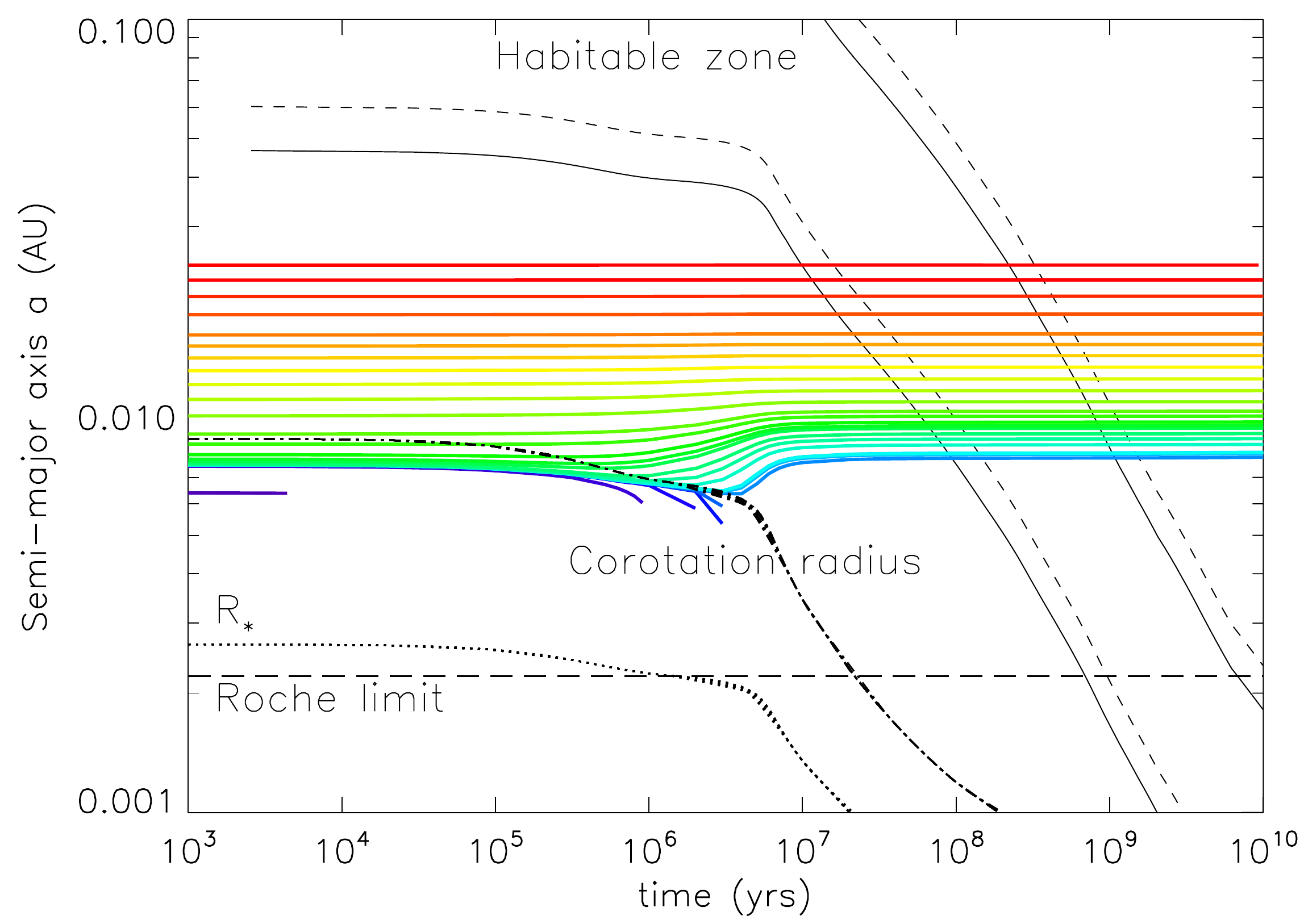}
	\caption{Evolution of the semi-major axis of a $1~\Mearth$ planet orbiting a $0.04~\Msun$ BD. The solid colored lines are for different initial semi-major axis. The dashed dotted line represent the radius of the BD and the dotted line the corotation radius. The habitable zone is also plotted using two different conditions of habitability (solid line : the planet receives an incoming flux of $400$~W/m$^2$, dashed line : $240$~W/m$^2$, see section \ref{part3:1} for more details). The bold dashed line represents the Roche limit.}
	\label{aM004_Mp1_eo001}
	\end{center}
	\end{figure}

The range of evolutionary pathways for this example is shown in Figure \ref{aM004_Mp1_eo001}. As other parameters are held fixed, the evolution is determined by the initial position of the planet with respect to the corotation distance.  

If the planet's mean motion is slower than the BD's rotation rate -- ie if the planet's initial semi-major axis is larger than the corotation radius -- then the tidal bulge raised by the planet on the BD is in advance with respect to the position of the planet. The resulting effect of that misalignment is that the planet is accelerated and its semi-major axis increases. Eventually, as the radius of the BD decreases the tidal dissipation in the BD decreases and the planet evolution stops at a given semi-major axis. This final position depends on a number of parameters as will be presented in the next subsection.

On the other hand, if the planet's mean motion is faster than the BD's rotation rate -- ie if the initial semi-major axis is smaller than the corotation radius -- then the tidal bulge raised by the planet on the BD lags behind the position of the planet. The planet is therefore slowed down and its semi-major axis decreases. If the planet spirals in very quickly then it will directly hit the BD surface, but if its evolution is slower --  if it survives the first few million years -- then it will reach the Roche limit and be ripped apart. 
Such planetary destruction is likely to form a debris disk around the BD, which is potentially observable with next-generation instruments such as JWST.  

A planet's fate depends on the first few million years of evolution, and its semi-major axis is stabilized in about a hundred million years.

The initial orbital radius offers a first order approximation of planets' tidally-driven orbital evolution, but as the BD's spin rate increases in time some intermediate cases can be seen.
For example, if a planet's initial orbital distance is inside the corotation radius but sufficiently close to it, the planet may undergo a two-phase evolution.  First, the planet will begin falling onto the BD.  However, if the BD's rotation period decreases faster than the planet's orbit shrinks, then the corotation distance can pass by the planet's orbit.  Then, the direction of tidal evolution is reversed and the planet migrates outward. 

%\st{It will eventually reach the Roche limit where the tidal forces are so strong that the planet is no longer gravitationally bound and disintegrates. }
%
%\st{If a planet's initial semi-major axis is large enough, then it is pushed away by an amount that decreases with the initial orbital radius.  If its initial semi-major axis is too small, then it spirals inward to its doom; if the planet evolves rapidly it will directly fall onto the BD's surface but if its fall is slower -- i.e. if it survives the first few million years -- then it will reach the Roche limit and be ripped apart by the strength of tidal forces. 
%Such planetary destruction is likely to form a debris disk around the BD, which is potentially observable with next-generation instruments such as JWST. }
%
%\st{This will determine if the planet is going to fall towards the BD (crash : behavior $\mathcal{A}$) of pushed away by the tidal forces (outward migration : behavior $\mathcal{B}$).}  

	\begin{figure}[h!]
	\begin{center}
	\includegraphics[width=8cm]{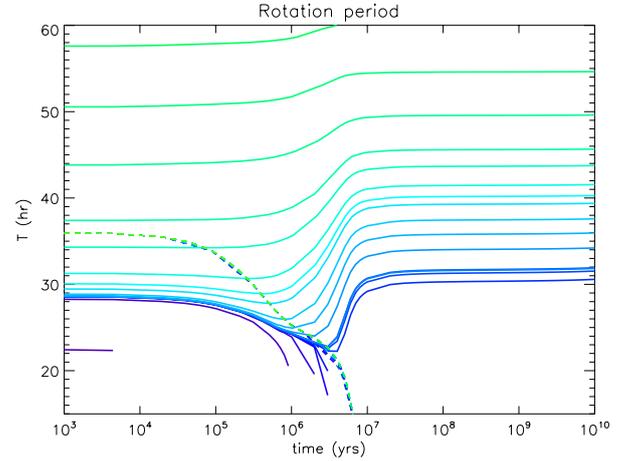}
	\caption{Evolution of the rotation period of a $1~\Mearth$ planet evolving around a $0.04~\Msun$ BD. The full lines represent the evolution for different initial semi-major axis. The dashed lines represent the rotation period of the BD. }
	\label{aM004_Mp1_rotzoom}
	\end{center}
	\end{figure}

Figure \ref{aM004_Mp1_rotzoom} shows the evolution of the rotation periods of the BD and planet for the planets initially close to the corotation distance in Fig. \ref{aM004_Mp1_eo001}.  As described above, a planet that forms inside the corotation radius may survive tidal evolution. It begins to fall towards the BD and its rotation period decreases. However, the BD is contracting and its rotation period also decreases. Depending on the rate of decrease of the two quantities the BD rotation period may catch on the planet rotation and reverse the direction of the planet's tidal evolution. Once the rotation period of the planet becomes longer than the BD's, tidal forces push the planet away. This intermediate evolution is very rare because the corresponding range of initial semi-major axes is very narrow. In the case of Fig. \ref{aM004_Mp1_eo001} for a BD of  mass  $0.04~\Msun$ and a planet of $1~\Mearth$, the width of this range is only $\sim10^{-3}$~AU (from $\sim7.5$ to $8\times10^{-3}$~AU).

Fig \ref{aM004_Mp1_rotzoom} shows that, for the most extreme cases, the planet's influence acts to accelerate the BD just before crossing the corotation radius. For less extreme cases, the planet does not influence the BD rotation.

%In Fig. \ref{aM004_Mp1_rotzoom} is represented the evolution of the rotation periods of the two bodies. A planet which is inside the corotation radius may survive the tidal evolution. It begins to fall towards the BD so its rotation period decrease. However, the BD is speeding up and its rotation period also decreases. Depending on the rate of decrease of the two quantities the stellar rotation period may catch on the planet rotation and thus inverting the tidal effect on the planet. Once the rotation period of the planet becomes slightly bigger than the BD's, the tidal forces will push the planet away. We can see that for the most extreme cases, the planet succeeds in accelerating the BD a bit just before crossing the corotation radius. For less extreme cases, the planet does not influence notably the rotation of the BD.

 %%%%%%%%%%%%%%%%%%%%%%%%%%%%%%%%%%%%%%%%%%%%%%%%%%%%%%%%%%%%%%%
 
\subsection{Influence of parameters}

In this section we investigate the effect of each parameter while keeping the others fixed at arbitrarily-chosen default values. Table \ref{table} lists these default values. 

\begin{table}[h]
\begin{tabular}{|c|c|c|c|c|c|}
\hline
$t_0$ & $M_{BD}$ & $M_p$          & $\sigma_{BD}$                         & $\sigma_{p}$                            & $e_0$   \\
(Myr)  & ($\Msun$) & ($\Mearth$)  &  (kg$^{-1}$m$^{-2}\sec^{-1}$) & (kg$^{-1}$m$^{-2}\sec^{-1}$) &             \\
\hline
\hline
$1$   &  $0.04$      & $1$                  &  $2.006 \times 10^{-53}$         & $8.577 \times 10^{-43}$         & 0.01 \\
\hline
\end{tabular}
\caption{Table of default values of parameters.}
\label{table}
\end{table}

For every set of parameters several simulations were done with different initial semi-major axis.

%%%%%%%%%%%%%%%%%%%%%%%%%%%%%%%%

\subsubsection{Influence of initial eccentricity}

The initial orbital eccentricity can play a role in the early stages of tidal evolution. For an eccentric orbit, the tides raised by the BD in the planet are strong and dissipation in the planet is important. This leads to a quick decrease of the eccentricity to minimize the planet's internal stress. Once the eccentricity becomes negligible, the dissipation in the planet is also negligible and the evolution is governed by the tides created in the BD.

For a fixed semi-major axis, a higher eccentricity implies a faster planetary rotation because the planet synchronizes at periastron where the orbital velocity is larger \citep{Hut1981}.  Thus, the initial semi-major axis $a_{switch}$ that divides behaviors $\mathcal{A}$ (inward migration and crash) and $\mathcal{B}$ (Inward migration but survival of the planet or outward migration) is farther from the BD for larger initial eccentricity.

	\begin{figure}[h!]
	\begin{center}
	\includegraphics[width=8cm]{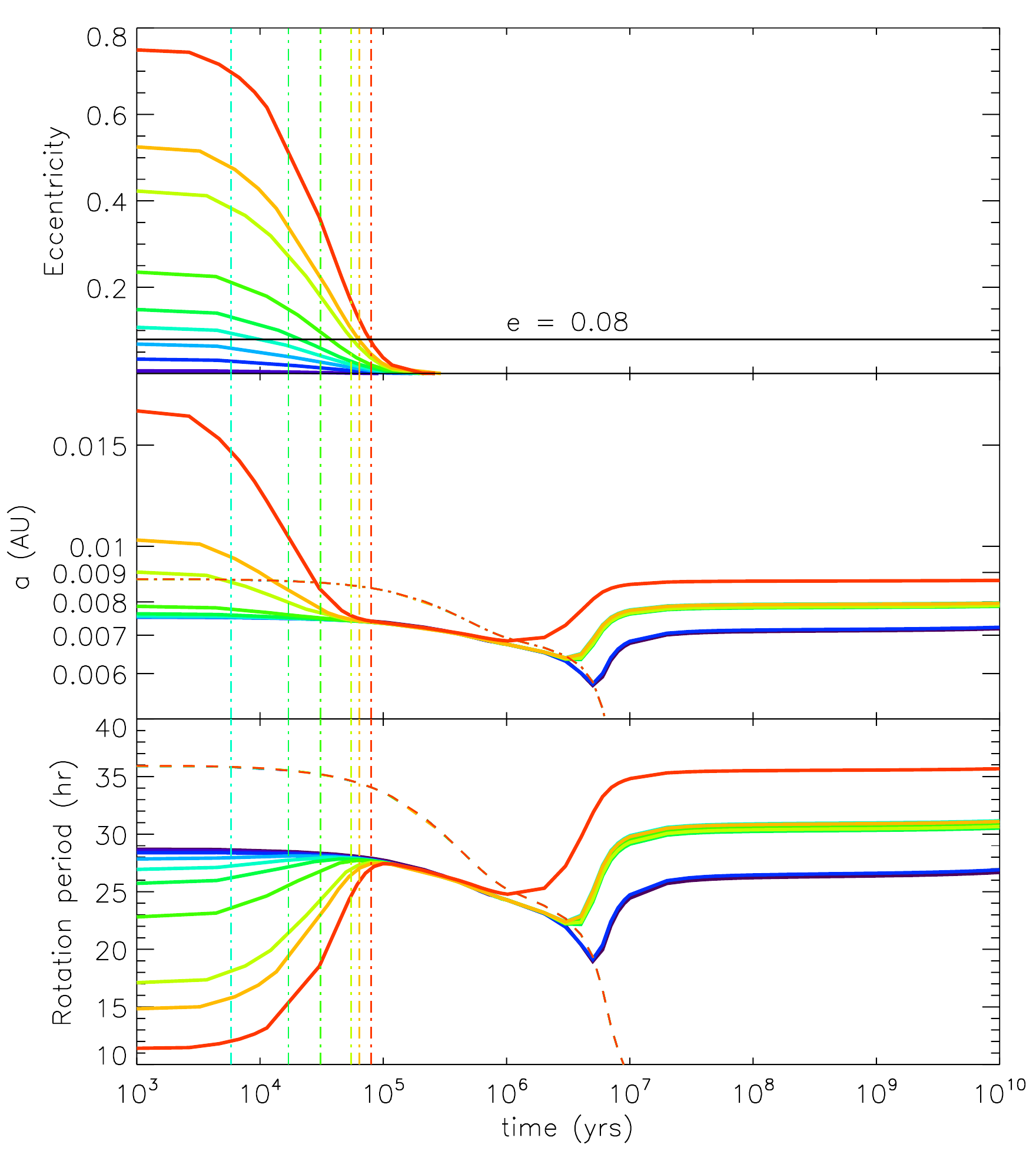}
	\caption{Evolution of semi-major axis, eccentricity and rotation period of a $1~\Mearth$ planet evolving around a $0.04~\Msun$ BD. Each solid line represents a planet with a different initial eccentricity. The initial semi-major axis for each case is  $a_{switch}$, which increases with eccentricity. The dashed-dotted line represent the corotation radius and the dashed line the rotation period of the BD. The vertical lines show the time at which the eccentricity effects become negligible for each initial eccentricity.}
	\label{aerotM004_Mp1_eo}
	\end{center}
	\end{figure}

Figure \ref{aerotM004_Mp1_eo} shows the evolution of a suite of planets starting at $a_{switch}$ with initial eccentricities of $0$, $0.01$, $0.05$, $0.1$, $0.15$, $0.2$, $0.3$, $0.5$, $0.6$ and $0.8$. The early evolution depends strongly on the initial eccentricity.  At early times an eccentric planet rotates faster than the BD such that the planet's semimajor axis decreases and its eccentricity is damped.  The planet's orbit shrinks but, due to the eccentricity-dependence of pseudo-synchronization, the rotation slows.  Then, when the BD's rotation period drops below the planet's, the planet is pushed away.  It is interesting to note in Fig. \ref{aerotM004_Mp1_eo} that, for these cases which all started at $a_{switch}$, the evolution of the different cases converges once the eccentricity drops below $\sim0.08$.  

	\begin{figure}[h!]
	\begin{center}
	\includegraphics[width=7cm]{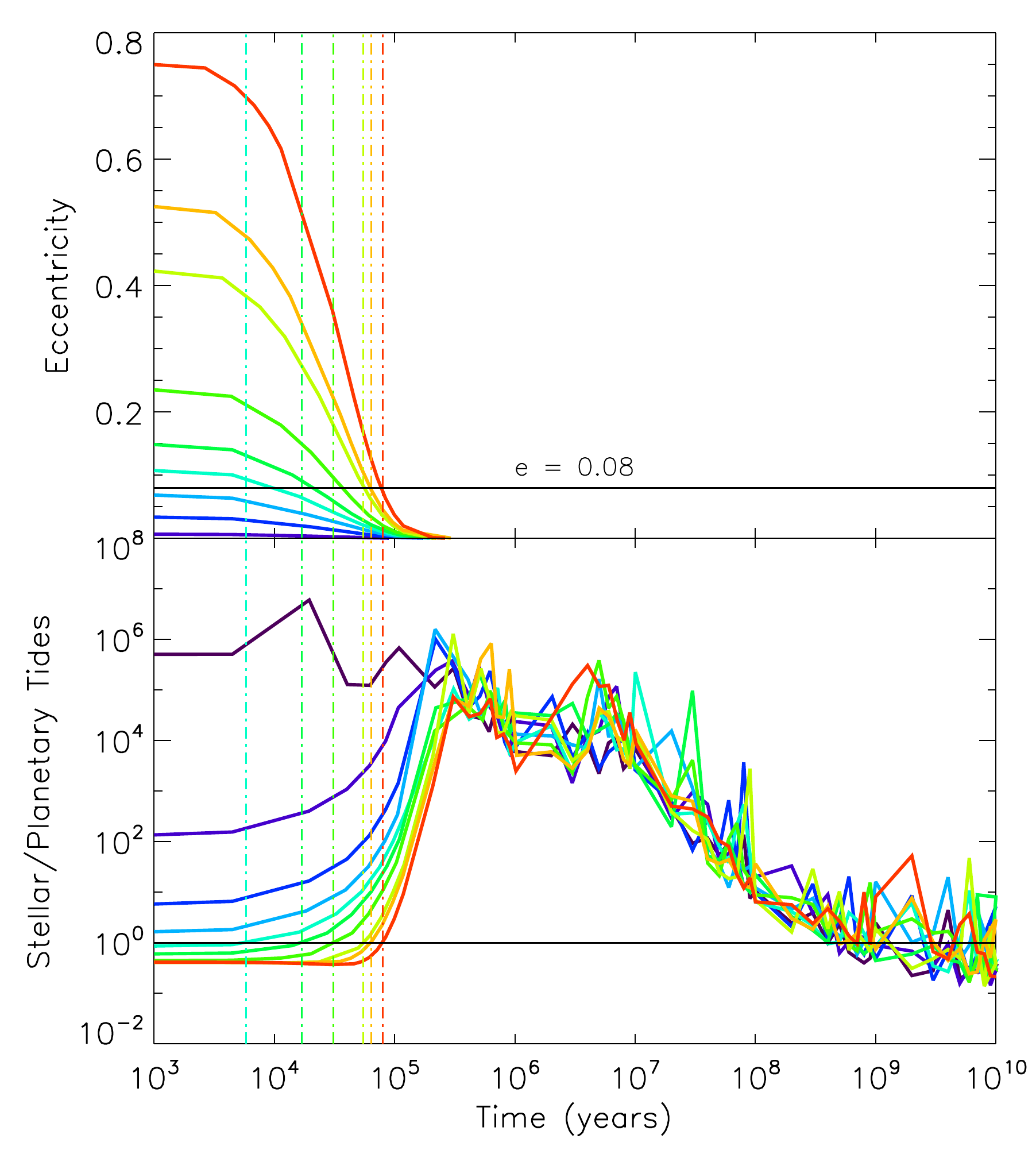}
	\caption{Evolution of eccentricity and of the ratio of planetary over BD tides of a $1~\Mearth$ planet evolving around a $0.04~\Msun$ BD. The full colored lines represent the evolution for different initial eccentricity. For a non zero eccentricity, the planetary tide dominates. When the eccentricity reaches $0.08$, the strength of the BD tide is equal to the strength of the planetary tide. The vertical lines show the time at which the BD tide starts to win over planetary tide.}
	\label{epsM004_Mp1_eo}
	\end{center}
	\end{figure}

For circular planetary orbits the only tidal bulge is the one raised on the BD by the planet -- referred to as the BD tide -- and the evolution is controlled by dissipation within the BD.  However, for eccentric orbits there is an additional bulge raised on the planet, called the planetary tide, and the evolution depends on dissipation within both the planet and BD. 
%
%For a $1~\Mearth$ planet and a $0.04~\Msun$ BD the ratio of the dissipation timescales is : $K_{\rm p}/K_\ast = \Ts/\Tp \sim 10$ at $t_0 = 10^6$~yr and $\sim 10^9$ at $1$~Gyr. The difference is due to the radius decrease of the BD, which makes its dissipation decrease - cf equation (\ref{Tp}).  {\bf SEAN: explain this more -- where is it coming from and exactly what depends on e (it doesn't look like Tp or Tstar do, right?)  I added the previous paragraph to motivate this but you should clarify and make sure it fits}

If the initial eccentricity is $0$, then equation (\ref{Hansena}) can be simplified as : 

\begin{equation}\label{Hansenasimp}
\frac{1}{a}\frac{\d a}{\d t} = - \frac{1}{T_{BD}}\Big[1-\frac{\Omega_{BD}}{n}\Big].
\end{equation}

If the orbit of the planet is circular, the planet is in perfect synchronization so $\Op = n$, and the first term of equation (\ref{Hansena}) is zero. As expected, only the tides raised by the planet in the BD remain. If $\Op = n > \Omega_{BD}$, $\frac{\d a}{\d t} < 0$ and the planet will fall towards the BD (behavior $\mathcal{A}$) . If $\Op = n < \Omega_{BD}$, $\frac{\d a}{\d t} > 0$ and the planet will be pushed away (behavior $\mathcal{B}$).  As discussed above, the fact that $\Omega_{BD}$ increases with time allows for the existence of the intermediate case in which the planet starts falling but is later pushed outward when $\Omega_{BD}$ becomes larger than $\Op=n$.

If the initial eccentricity is non zero, and for $\Omega_{BD} \sim\Op$ at time zero, the difference between the strength of BD and planetary tides comes from the dissipation timescales. When the eccentricity is large, the planetary tide -- i.e., the rate of energy dissipation within the planet -- dominates the evolution because the amplitude of the tidal bulge varies over the course of the planet's orbit. For a $1~\Mearth$ planet and a $0.04~\Msun$ BD the ratio of the dissipation timescales is : $T_{BD}/\Tp \sim 10$ at $t_0 = 10^6$~yr and $\sim 10^9$ at $1$~Gyr. Rocky planets dissipate more than BDs and the difference in time is due to the shrinking radius of the BD, which makes its dissipation decrease -  see equation (\ref{Tp}). So while the eccentricity is non zero, the tides raised on the planet by the BD will prevail. In this case, $Na1(e) \gg Na2(e)$, so $\frac{\d a}{\d t}$ is negative and the planet is pulled inward. This is always true because for a planet in pseudo-synchronization the planetary tide always acts to decrease the orbital distance \citep[see][]{Leconte2010}. But when the eccentricity is small the BD tide dominates and determines the evolution of the planet: if the planet is beyond the corotation radius, $\frac{\d a}{\d t}$ is positive so the planet is pushed outward, and if it is inside, $\frac{\d a}{\d t}$ is negative so the planet is pulled inward. 
 
Figure \ref{epsM004_Mp1_eo} shows the ratio of the strength of BD to planetary tides for the case shown in Fig. \ref{aerotM004_Mp1_eo}, where the strength of each tide is the absolute value of its contribution to $\frac{\d a}{\d t}$. While $e \gtrsim 0.08$, the planetary tide dominates but when $e \lesssim 0.08$ the BD tide dominates. This is true for all cases studied and analytically we can show that for an eccentricity of this order of magnitude BD and planetary tides have the same strength for the orbital distance evolution. At early times the planetary tide acts to damp the orbital eccentricity and shrink the planet's orbit but when the eccentricity drops below $\sim0.08$ the BD tide takes over and the orbit expands.   At this point the planet has a constant deformation and the ratio of the two tides is very high ($\gtrsim 10^4$) because the planetary tide is so small. As the planet is pushed outward the BD contracts and the BD tide also decreases to zero.

%%%%%%%%%%%%%%%%%%%%%%%%%%%%%%%%

\subsubsection{Influence of the BD mass}
\label{massstar}

The BD mass is important for tidal evolution because the more massive the BD the stronger the tides raised on the planet. The BD mass affects its response to the planet's gravity, i.e., the tides that the planet raises on the BD (referred to as the BD tides).

The dissipation timescale for the BD is simply calculated by switching the $p$ and $\ast$ indices in Eqn (\ref{Tp})  -- and replacing $\ast$ by BD -- to yield:

\begin{equation}
T_{BD} = \frac{1}{9}\frac{M_{BD}}{\Mp(\Mp+M_{BD})}\frac{a^8}{R_{BD}^{10}}\frac{1}{\sigma_{BD}}
\end{equation}

The BD mass has an indirect influence on $T_{BD}$ via the BD radius $R_{BD}$. More massive BDs have larger radii -- for  ages younger than $10^8$~yrs -- and thus heavier BDs have a shorter dissipation timescales, which means that tides are stronger. 

The BD mass also has an influence on its rotation, as seen in Fig. \ref{rotperiodBD_article}. The initial conditions taken in this study are that heavier BDs rotate slower, which is consistent with observations \citep{Herbst2007}. The corotation radius is therefore farther from heavier BDs, and they rotate slower so the tides will be less efficient than for lighter BDs, which have smaller corotation radius and higher rotation velocities.

	\begin{figure}[h!]
	\begin{center}
	\includegraphics[width=7cm]{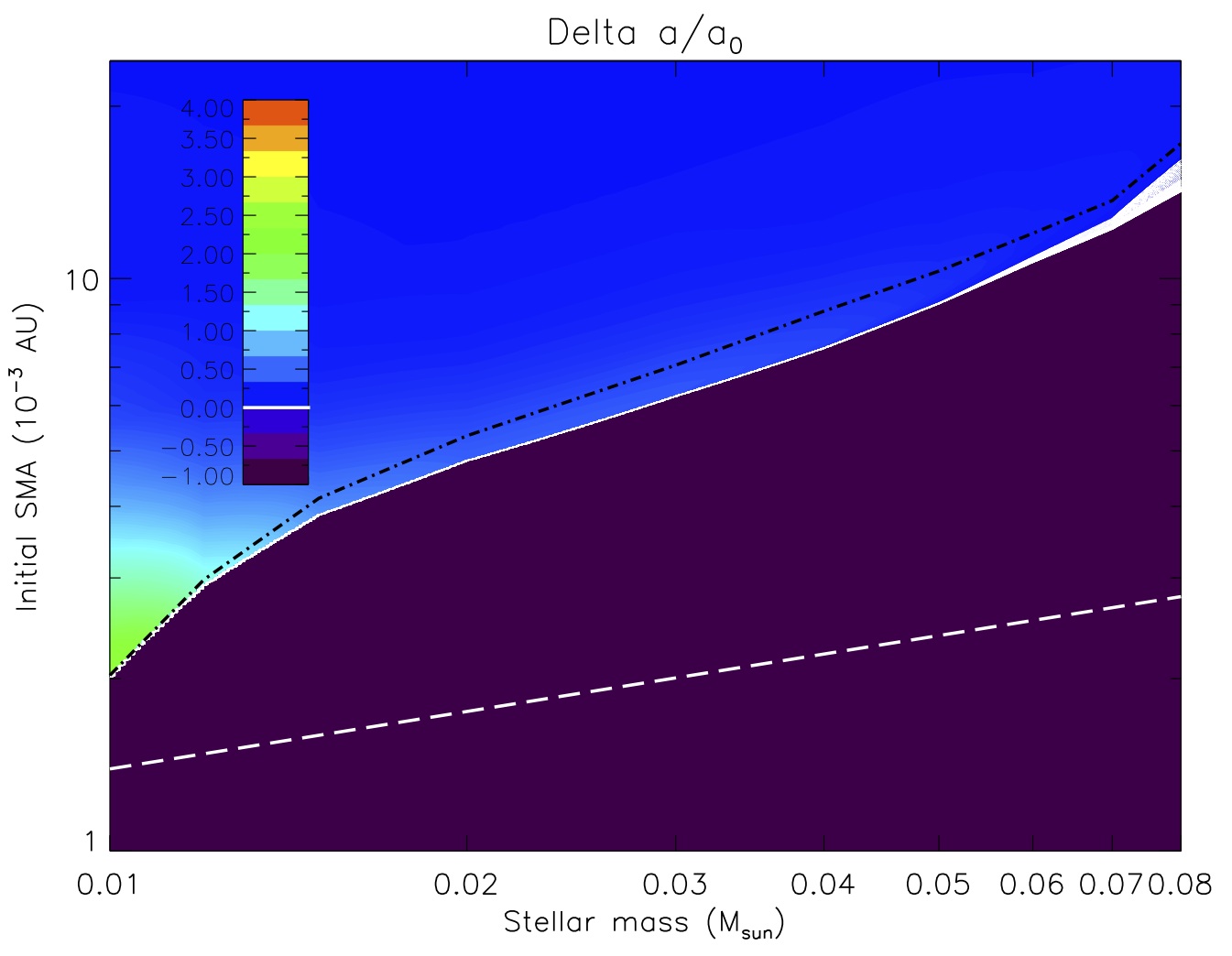}
	\caption{Map of $\Delta a/a_0$ for a range of initial semi-major axis and different BD masses. Time zero is $10^6$~yrs. The deep purple zone is the region for which the planets don't survive and the white zone is where the planets approach the BD without falling on it. The dashed dotted line is the corotation distance at the beginning of integration and the dashed line is the Roche limit.}
	\label{supermapp106}
	\end{center}
	\end{figure}
	
%{\bf SEAN: why not put the fig with time zero = 1 Myr here?  It seems better to show the good results first instead of the weird one with unphysical rotation rates first.  You can still refer to the fig when you use the longer time zero in the later section.  In fact, you could make this next part a separate subsection called "Influence of the initial BD spin rate" and put it after the "time zero" subsection for continuity.}

This dependance of the BD mass can be seen in Fig. \ref{supermapp106} and in Fig. \ref{supermapp107} in the next subsection \ref{inftime}.

Here, we define the quantity $\Delta a/a_0$, the fractional change in semi-major axis over the planet's evolution, as $\Delta a/a_0 = (a_f-a_0)/a_0$, where $a_0$ and $a_f$ are the semi-major axis at $t = t_0$ and at $t = t_f = 10^{10}$~yrs, respectively. When this quantity is positive, the planet is pushed away and when it is negative it comes closer to the BD.  When a planet falls onto the BD, $\Delta a/a_0 = -1$.

In Figure \ref{supermapp106}, the dark zone is the region for which the planet falls on the BD, the white zone is the region for which the planet approaches the BD without falling on it. This last situation happens for planets inside the corotation radius which are saved by the acceleration of the BD, however the BD tide is not strong enough to push them back farther than their initial position. The zone between the corotation line and the white zone corresponds to planets that begin falling but are then pushed farther than their initial position.

\subsubsection{Influence of the initial BD spin rate}

Given that there remains considerable uncertainty in the spin evolution of BDs, we performed a suite of tidal evolution calculations assuming that all BDs start with the same initial rotation period of 36 hours.  This exercise isolates the dependence on the BD mass, although in a regime that is probably not realistic given current observations \citep{Herbst2007}.  In addition, we did not take care in these integrations to avoid very fast final rotation periods that might be non physical because they may approach breakup velocity.

	\begin{figure}[h!]
	\begin{center}
	\includegraphics[width=7cm]{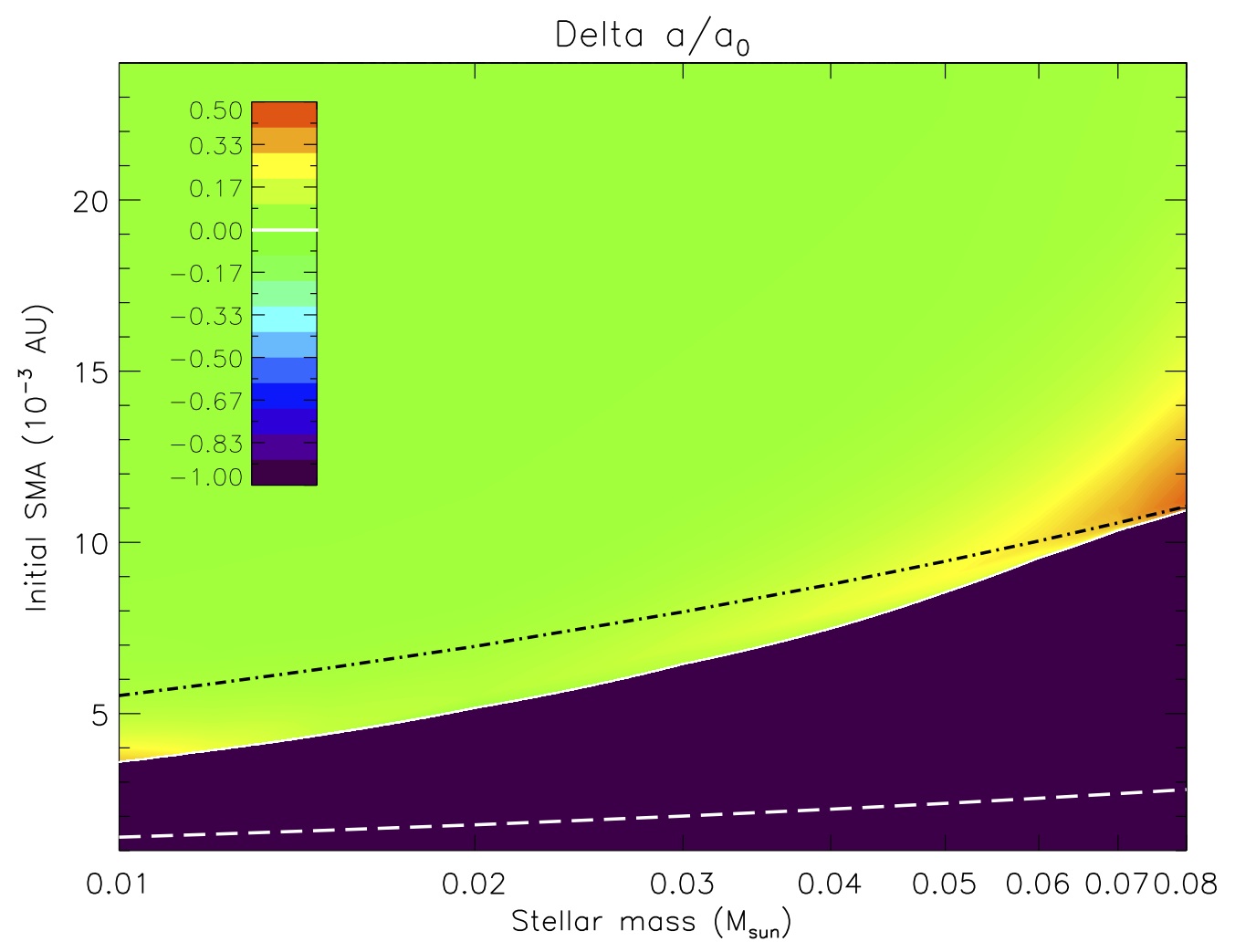}
	\caption{Map of $\Delta a/a_0$ for a range of initial semi-major axis and different BD masses. Each BD has the same initial rotation velocity. The dashed dotted line is the corotation distance at the beginning of integration and the dashed line is the Roche limit.}
	\label{supermap106_Ms_36hr}
	\end{center}
	\end{figure}

Figure \ref{supermap106_Ms_36hr} shows a map of $\Delta a/a_0$ as a function of the initial semimajor axis and BD mass.

In contrast with Fig. \ref{supermapp106} of the previous subsection, this map shows that the planets inside the corotation distance are efficiently pushed away. The initial corotation distance scales here as $(\Ms+\Mp)^{1/3}\sim\Ms^{1/3}$ as we got rid of the initial BD rotation dependance. 

%The result can be seen in Fig. \ref{supermap106_Ms_36hr}, where a map of $\Delta a/a_0$ as a function of the initial semimajor axis and BD mass is presented. The dark zone is the region for which the planet falls on the BD, the white zone is the region for which the planet approaches the BD without falling on it. This last situation happens for planets inside the corotation radius which are saved by the acceleration of the BD, however the BD tide is not strong enough to push them back farther than their initial position. The zone between the corotation line and the white zone corresponds to planets, that begin falling but then that are pushed farther than their initial position. In contrast with Fig. \ref{supermapp106} of the next subsection, this map shows that the planets inside the corotation distance are efficiently pushed away. The initial corotation distance scales here as $(\Ms+\Mp)^{1/3}\sim\Ms^{1/3}$ as we got rid of the initial BD rotation dependance. 

Within this set of calculations with mass-independent initial spin rates, it is interesting to note that the planets that evolve according to the intermediate case --  planets that start interior to corotation but survive -- are more common around lower mass BDs.  This is because of the combination of the small masses and small radii of low mass BDs: planets inside the corotation radius fall slowly and thus have a longer time window for the corotation radius to catch up with them. This tendency is absent in Fig. \ref{supermapp106} because the different initial rotation velocities of the BD cancels this effect.

Furthermore, with identical initial spin rates low- and high-mass BDs push the planets farther than medium mass BDs. It is logical that higher mass BDs being heavier and bigger push planets farther but for low mass BDs the explanation is not as straightforward. The reason for this is that the low mass BDs evolve less than high mass BDs and they spin up relatively slowly because their radii do not decrease much. Surviving planets remain close to the BD for a long time so the tides remain active and keep on pushing the planets outward on timescales of $10$~Gyrs.

%%%%%%%%%%%%%%%%%%%%%%%%%%%%%%%%

\subsubsection{Influence of the time zero}
\label{inftime}

Gaseous protoplanetary disks are observed to dissipate after a few Myr around young stars \citep{Haisch2001, Hillenbrand2008}.  Disks around low-mass stars have systematically longer lifetimes, of up to roughly $10$~Myr \citep{Pascucci2009}.  The evolution of discs around BDs is not well constrained but it is reasonable to assume lifetimes between one and ten million years. If planets form at the same approximate temperature around BDs as around main sequence stars then their formation times should be shorter by a factor of $40$-$1000$, since the accretion time scales inversely with the orbital frequency (see \citet{Safronov1969}; for a comparison of simulated and calculated accretion times around low-mass stars see \citet{Raymond2007}).  Thus, planets around BDs probably form before the disk dissipates and so we consider "time zero" for our tidal calculations to be the time of the dissipation of the gaseous protoplanetary disk. 

We now compare the tidal evolution of planets with time zero of $1$ and $10$~Myr, which should bracket the range of reasonable values based on the arguments presented above.  The only important difference between these two initial conditions is between one and ten million years the BD contracts and spins up considerably.  Thus, at $10$~Myr the corotation radius is smaller (see Fig. \ref{rotperiodBD_article}).  

We performed two suites of tidal calculations of $1 \Mearth$ planets evolving around BDs with masses from $0.01-0.08 M_\odot$ for time zero values of $1$ and $10$~Myr.

\begin{figure}[h!]
	\begin{center}
	\includegraphics[width=7cm]{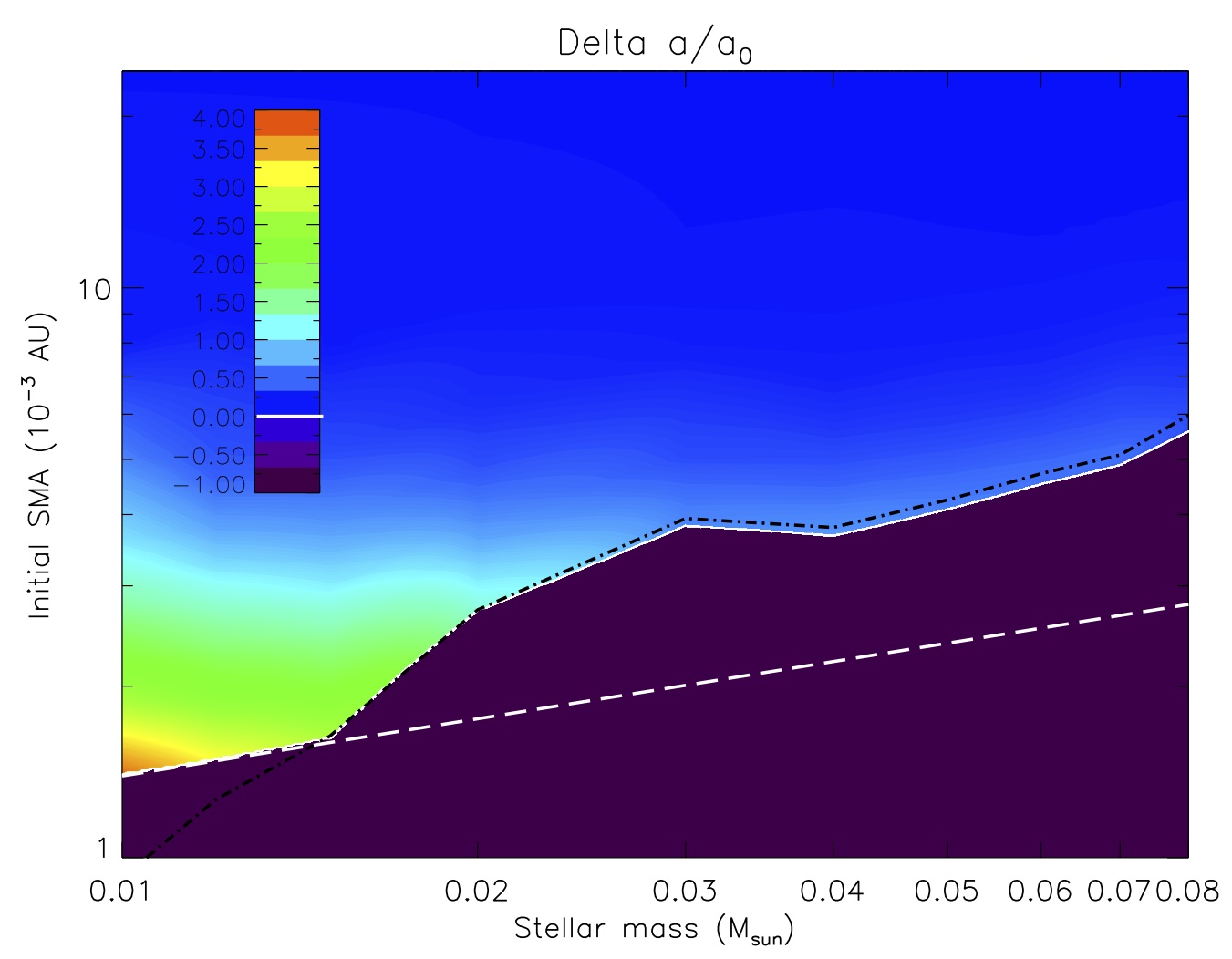}
	\caption{Map of $\Delta a/a_0$ for a range of initial semi-major axis and different BD masses. Time zero is $10^7$~yrs. The purple zone has shrunk compared to Fig. \ref{supermapp106} because the BD has spun up and the corotation radius is thus smaller.}
	\label{supermapp107}
	\end{center}
	\end{figure}

%	\begin{figure}[h!]
%	\begin{center}
%	\includegraphics[width=7cm]{supermap106logb.pdf}
%	\caption{Map of $\Delta a/a_0$ for a range of initial semi-major axis and different BD masses. The planetary formation time is $10^6$~yrs. The deep purple zone is the region for which the planets don't survive and the white zone is where the planets approach the BD without falling on it. The dashed dotted line is the corotation distance at the beginning of integration and the dashed line is the Roche limit.}
%	\label{supermapp106}
%	\end{center}
%	\end{figure}
%	
	Maps of $\Delta a/a_0$ as a function of the initial semimajor axis and BD mass are presented in Figures \ref{supermapp106} (for a time zero of $1$~Myr) and \ref{supermapp107} (for a time zero of $10$~Myr). 

The corotation distance varies as $\Omega_{BD}^{-2/3}(M_{BD}+\Mp)^{1/3}\sim\Omega_{BD}^{-2/3}M_{BD}^{1/3}$  such that for our chosen initial BD rotation rates (see Fig. \ref{rotperiodBD_article}) the corotation distance  increases  with increasing BD mass. Consequently, the range of semi-major axis resulting on surviving planets decreases with mass.  High-mass BDs cannot have planets as close-in as low-mass BDs. We can thus infer that there is a forbidden region larger than the one delimited by the Roche limit where no planet of $1 \Mearth$ should survive. This rough limit may be useful for future observations.

Figures \ref{supermapp106} and \ref{supermapp107} also show that planets orbiting low-mass BDs are pushed farther than those orbiting high mass BDs. This is due to the smaller initial rotation period of low-mass BDs, they compensate for their smaller masses by being fast rotators. From these maps, we can also observe that $\Delta a/a_0$ decreases to $0$ as the initial orbital distance increases, which is logical because tidal effects decrease quickly with distance.

After $1$~Myr the BD's rotation period is $36$~hours and by $10$~Myr it decreases to about $10$~hours.
During the interval from $1-10$~Myr the corotation radius has shrunk correspondingly so there is a wider range of initial semi-major axes that allows the planets to survive tidal evolution,  which may be advantageous for planet detection (see Discussion in Section \ref{Discussion}). 

The corotation distance does not have quite the same shape at $1$ and $10$~Myr (see Figs \ref{supermapp106} and \ref{supermapp107}) because of the dependance on $\Omega_{BD}$. Indeed, we have seen that the time evolution of the spin rate varies with the BD mass (Section 2.3.3). Fig. \ref{rotperiodBD_article} shows that the BD rotation period no longer increases monotonically with BD mass after $10$~Myr, especially for BD masses larger than $0.03~\Msun$.
This explains the decrease in the corotation distance between $M_{BD} = 0.03~\Msun$ and $0.04~\Msun$ in Fig. \ref{supermapp107}. For BDs with masses below $0.16 \Msun$, the Roche limit is outside the corotation distance so some planets exterior to corotation do not survive the evolution. As shown in \ref{supermapp107}, the forbidden region is defined as the maximum of the Roche limit and the corotation distance. In contrast with a time zero of $1$~Myr (Fig \ref{supermapp106}), here the corotation distance is coincident with the outer limit of the crash region. 

For $t_0 = 10^7$~yrs the BD rotates faster than at $t_0 = 10^6$~yrs so planets that survive are pushed away more strongly and the white zone where planets approach the BD without falling on it is even narrower.

%%%%%%%%%%%%%%%%%%%%%%%%%%%%%%%%

\subsubsection{Influence of planetary mass}
\label{infplanetmass}

Tidal effects are stronger for more massive bodies, so if the planetary mass is increased the planets are pushed farther or fall quicker. Moreover, the corotation radius increases as $(M_{BD}+\Mp)^{1/3}$ so heavier planets have slightly higher $a_{switch}$.

	\begin{figure}[h!]
	\begin{center}
	\includegraphics[width=9cm]{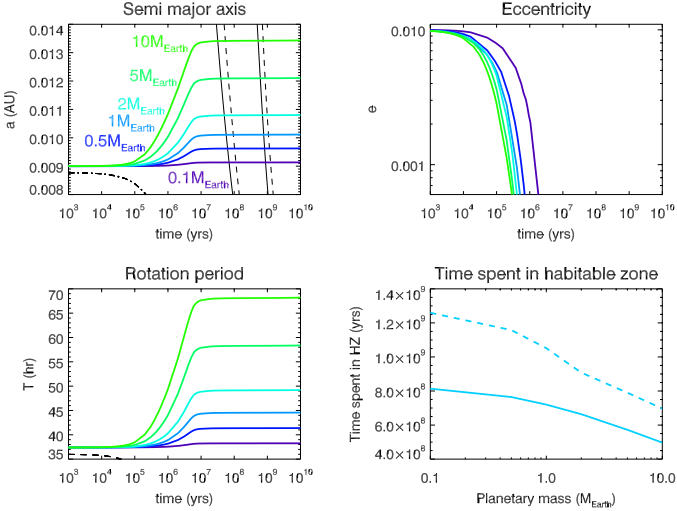}
	\caption{{\bf Tidal} evolution of planets with different masses starting at $9\times10^{-4}$~AU around a $0.04\Msun$ BD. The heavier planets are pushed away farther. The planetary masses are : $0.1$, $0.5$, $1$, $2$, $5$, $10~\Mearth$.}
	\label{aerotM004_eo001_Mp}
	\end{center}
	\end{figure}

Figure \ref{aerotM004_eo001_Mp} shows a set of tidal evolution calculations for planets with masses from 0.1-10 $\Mearth$, starting from just exterior to the corotation radius. For a planet of $0.1~\Mearth$ tides only have the effect of reducing the eccentricity, but there is no semi-major axis change. The planet mass has little effect on the eccentricity evolution; in all cases $e$ is damped to zero on a $1-10$~Myr timescale.  However, the final semi-major axis depends strongly on the planet mass. This is because, especially at early times before the BD contracts, the BD tide is stronger for larger planet masses and this is what drives the orbital expansion. 

For completeness, we ran a limited set of tidal calculations with planet masses of one Saturn mass and one Jupiter mass.

The planet's dissipation factor was taken to be the same one as the BD -- so the same as \citet{Hansen2010}'s $\sigma_p$. What is interesting for the case of gas giant planets is that they can have a significant influence on the tidal evolution of the BD. When the massive planet is close, angular momentum is transferred from the BD's rotation to the planet's orbit much more efficiently than for terrestrial planets.

	\begin{figure}[h!]
	\begin{center}
	\includegraphics[width=9cm]{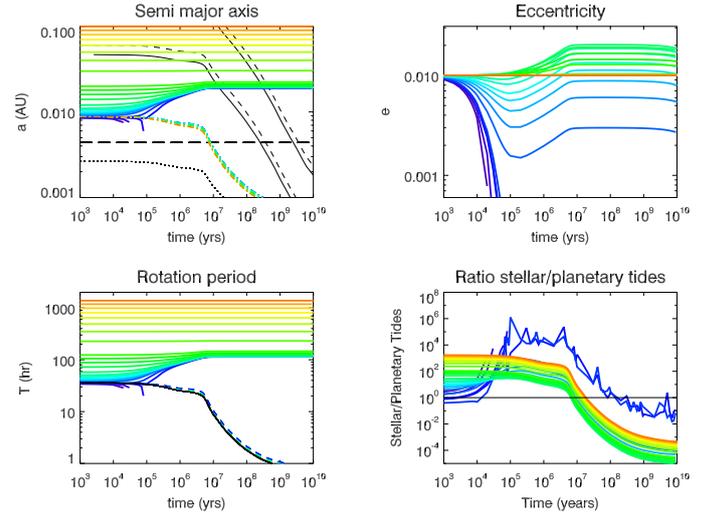}
	\caption{Tidal evolution of a Saturn mass planet starting at different initial semi-major axis around a $0.04~\Msun$ BD. The first panel shows the evolution of the semi-major axis (bold solid lines). The dashed dotted line represent the radius of the BD, the dotted line the corotation radius and the bold dashed line the Roche limit. The habitable zone is also plotted. The second panel shows the evolution of the eccentricity. The third panel shows the evolution of the rotation period of planets (solid lines) and the BD (dashed lines). The solid black line represents the rotation period of the BD if its evolution is governed only by conservation of angular momentum. The fourth panel shows the ratio of stellar over planetary tides and the horizontal line is a ratio of $1$.}
	\label{aerotM004_eo001_Msat}
	\end{center}
	\end{figure}

	\begin{figure}[h!]
	\begin{center}
	\includegraphics[width=9cm]{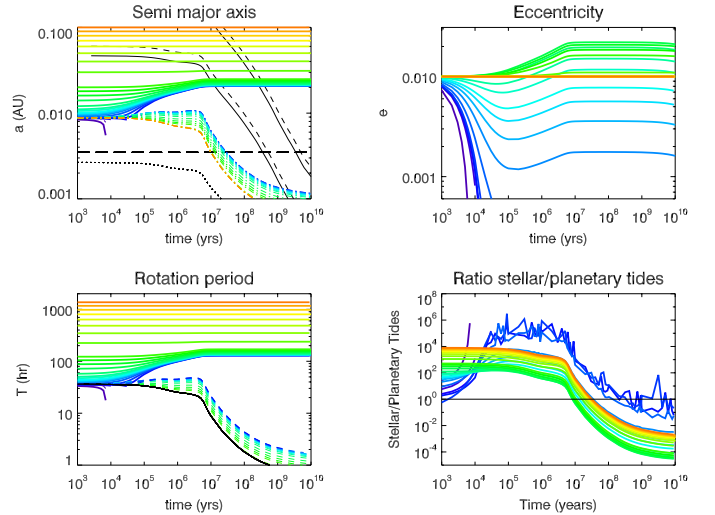}
	\caption{Tidal evolution of a Jupiter mass planet starting at different initial semi-major axis around a $0.04~\Msun$ BD. The third panel shows that a close-in Jupiter mass planet significantly slows the BD with respect to the rotation rate it would have if there was conservation of angular momentum.}
	\label{aerotM004_eo001_Mjup}
	\end{center}
	\end{figure}
	
Figures \ref{aerotM004_eo001_Msat} and \ref{aerotM004_eo001_Mjup} show the evolution of the gas giant's semi-major axis, eccentricity, and rotation period and the ratio of BD to planetary tides for a Saturn- and Jupiter-mass planet. The semimajor axis evolution for gas giants is qualitatively similar to the evolution for lower-mass planets, with larger-amplitude outward migration due to the planets' larger masses.  But this is where the similarities end.

\begin{figure}[h!]
	\begin{center}
	\includegraphics[width=7cm]{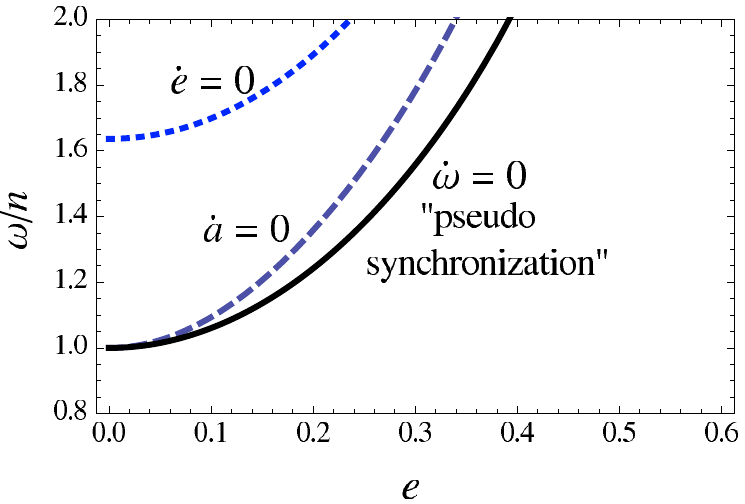}
	\caption{$\omega/n$ vs $e$ phase space, reproduced from \citet{Leconte2010} Fig. 1. Here, $\omega=\Omega_{p}$ or $\Omega_{BD}$. The planets are in pseudo-synchronization so $\Omega_{p}/n$ remains on the black solid line, thus the planetary tide always contribute in making the semi-major axis and eccentricity decrease. If the eccentricity is $\sim 0$, we can see that if $\Omega_{BD}/n >1$ the BD tide contributes in pushing planets away and if $\Omega_{BD}/n>18/11\sim1.64$, the eccentricity increases.}
	\label{phasediag}
	\end{center}
	\end{figure}

The eccentricity evolution is dramatically different for gas giants: the planetary tide still acts to decrease the eccentricity but the BD tide can act to increase it.  The phase diagram in Fig. 1. of \citet{Leconte2010} (reproduced in Fig. \ref{phasediag}) shows that for high $\Omega_{BD}/n$ the BD tide can indeed increase the eccentricity. This can be interpreted as a slingshot effect: the planet is kicked when approaching periastron, and if the kick is strong enough the trajectory gets more and more eccentric \citep{Hut1981}. 

At early ages the planets closest to the BD have a high mean motion and the BD has not accelerated yet so $\Omega_{BD}/n<\frac{18}{11}\frac{Ne1(e)}{Ne2(e)}$. This criterion comes from the BD tide contribution in Eqn. \ref{Hansene} and defines the limit $\dot{e}=0$ in Fig. 12 (reproduced from \citet{Leconte2010}). For $\Omega_{BD}/n<\frac{18}{11}\frac{Ne1(e)}{Ne2(e)}$, the BD tide acts to decrease the eccentricity. So for close planets at early ages, both BD and planetary tides act to decrease the eccentricity. This effect is visible in Figures \ref{aerotM004_eo001_Msat} and \ref{aerotM004_eo001_Mjup} for ages between $10^3$~yrs and $10^5$~yrs. When the BD spins up, the criterion $\Omega_{BD}/n<\frac{18}{11}\frac{Ne1(e)}{Ne2(e)}$ is no longer met so the BD tide acts to increase the eccentricity -- this is the slingshot effect. The planetary tide is not strong enough to counteract the eccentricity pumping from the BD. Figures \ref{aerotM004_eo001_Msat} and \ref{aerotM004_eo001_Mjup} show this phase between the ages of $10^5$~yrs and $10^7$~yrs. As the BD radius decreases, the BD tide weakens and after a few Gyr, the planetary tide becomes dominant and acts to once again decrease the eccentricity.

Planets that start farther from the BD have slower mean motion so $\Omega_{BD}/n>\frac{18}{11}\frac{Ne1(e)}{Ne2(e)}$ and the BD tide acts to increase the eccentricity. These planets skip the first stage of the evolution previously described.  Their eccentricity is pumped by the BD and when the BD tide gets weaker than the planetary tide the eccentricity decreases. For planets farther than $0.04$~AU, both tides are negligible and there is no effect on the semi-major axis or eccentricity.

\begin{figure}[h!]
	\begin{center}
	\includegraphics[width=9cm]{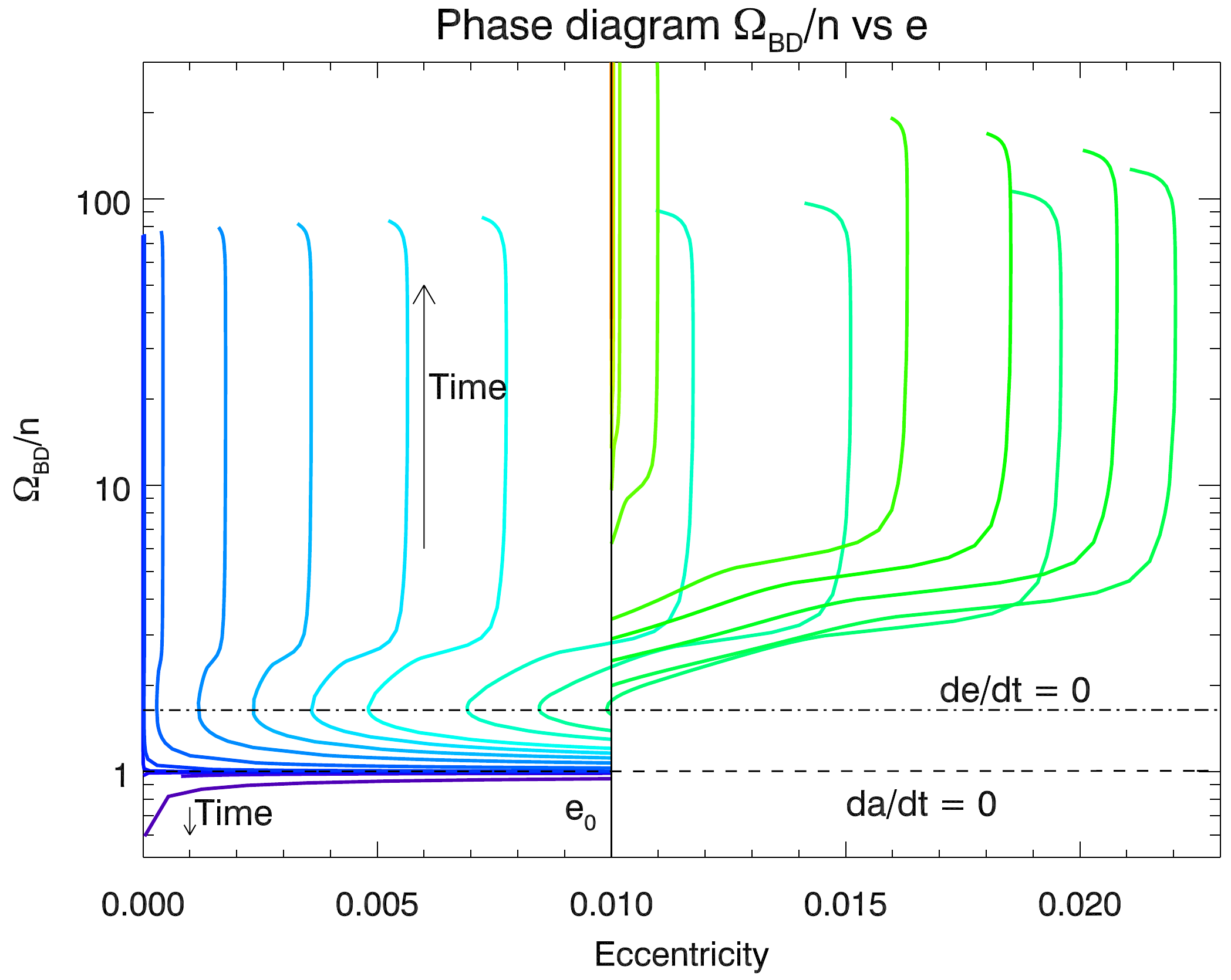}
	\caption{Evolutionary tracks of Jupiter mass planets orbiting a $0.04~\Msun$ BD. The evolution starts on the solid black line of $e_0=0.01$. The different solid lines are for different initial semi-major axis. The dashed line is where $\dot{a}=0$ and the dashed dotted line is where $\dot{e}=0$. The quantities $\dot{a}$ and $\dot{e}$ are positive above the zero value line.}
	\label{phasediagJUP}
	\end{center}
	\end{figure}

Figure \ref{phasediagJUP} shows the evolutionary tracks of Jupiter mass planets beginning at different semi-major axis in the phase diagram $\Omega/n$ vs eccentricity.  It shows the different phases of the tidal evolution as explained in the previous paragraph depending if $\Omega_{BD}/n$ is bigger or smaller than $\frac{18}{11}\frac{Ne1(e)}{Ne2(e)}$.

In this case, as can be seen in the fourth panel of Fig. \ref{aerotM004_eo001_Mjup}, the BD tide dominates the evolution. So Fig. \ref{phasediagJUP} shows only the effect of the BD tide on the evolution of the system except at the very end of the evolution when the planetary tide is becoming dominant. This is why there is a decrease in eccentricity in the $\dot{e}>0$ region at the end of the evolutionary tracks.

The key difference in evolution for gas giants come from the ability of massive planets to affect the BD's rotation such  that the BD rotation period can no longer be approximated by the simple conservation of angular momentum during the BD's contraction. Gas giants are so massive that they act to reduce the acceleration of the BD significantly during contraction. Planets outside the corotation radius have slower mean motion than the BD's rotation rate, so the system tends to transfer angular momentum from the BD to the planet making the BD slow down and the planet spin up and therefore move outward. The stellar tide efficiently pushes the planets away. The efficiency of the spinning down of the BD is measured with $f_{tides}$ (see Equation \ref{Hansenosftides}). This parameter is negligible for terrestrial planets but for a Jupiter mass planet $T_{BD}$ is smaller so $f_{tides}$ is larger. Fig. \ref{aerotM004_eo001_Mjup} shows that for close-in Jupiter mass planets the rotation rate of the BD does not quite follow the evolution governed by the conservation of angular momentum. As in the first few million years the BD does not contract much, the effect is visible, but at $t\sim10^7$~yrs the contraction of the BD becomes significant and the effect is overpowered and the rotation rate of the BD increases accordingly to the conservation of angular momentum.

%%%%%%%%%%%%%%%%%%%%%%%%%%%%%%%%

\subsubsection{Influence of BD dissipation}
\label{Infstdiss}

The BD dissipation is a poorly-constrained parameter.  Here we adopt Hansen's (2010) value for BDs and assume that it remains constant in time and constant for all BD masses. We note that \citet{Hansen2010} considers that the dissipation factor for gas giants is approximatively the same as for BDs. Indeed, gas giants and BDs are thought to have similar internal structures \citet{Baraffe2003}. As the BD can have a low rotation period it is interesting to compare the value of the dissipation factor we chose in comparison to a low rotation period gas giant dissipation factor. \cite{Leconte2010} gives a estimation of the dissipation in Jupiter $k_{2,J}\Delta T_{J} \lesssim 2-3\times 10^{-2}$~s, which corresponds to $\sigma_J \sim 6\times 10^{-52}~\kg^{-1}\m^{-2}\sec^{-1}$. Our dissipation factor being $\sigma_{BD} = 2.006 \times 10^{-53}~\kg^{-1}\m^{-2}\sec^{-1}$, the parameter space exploration includes BDs which have the same dissipation factor as Jupiter.

To test the effect of the BD dissipation on the tidal evolution, we performed a set of tidal calculations varying the dissipation factor $\sigma_{BD}$ by several orders of magnitude.  In each integration, the BD mass was $0.04~\Msun$, the planet mass was $1~\Mearth$, and the planet started just exterior to the corotation radius.  

\begin{figure}[h!]
	\begin{center}
	\includegraphics[width=9cm]{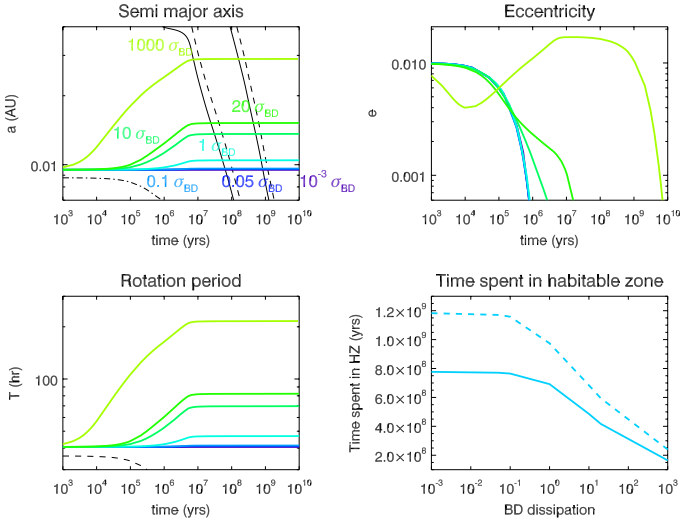}
	\caption{Tidal evolution of a $1~\Mearth$ planet starting at $9.5\times10^{-4}$~AU around a $0.04~\Msun$ BD for different BD dissipation. The planet is pushed farther if the BD dissipation is higher. The BD dissipation factors are : $0.001$, $0.05$, $0.1$, $1$, $10$, $20$, $1000\times\sigma_{BD}$.}
	\label{aerotM004_Mp1_eo001_ss}
	\end{center}
	\end{figure}

Figure \ref{aerotM004_Mp1_eo001_ss} shows the results of tidal calculations varying $\sigma_{BD}$ from $0.001$ to $1000$ times our fiducial value. 

The higher the BD dissipation factor, the stronger the tides created by the planet in the BD, so the farther the planet is pushed. For BD dissipation factors of $0.05\times\sigma_{BD}$ or smaller, tides are not strong enough to significantly change the planet's orbit.  
%{\bf However, the eccentricity evolution does not depend on the BD dissipation factor.  For BD dissipation rates (roughly $10\times\sigma_{BD}$ or more), the planet is quickly pushed to larger semimajor axis where the eccentricity damping is slower but this is a relatively weak effect.  }
The eccentricity evolution depends on the BD dissipation factor. For a low dissipation factor, the eccentricity evolution is governed by the planetary tide and decreases to zero. For a high BD dissipation factor -  for dissipation factor $\gtrsim 10~\sigma_{BD}$ - a significant change can be seen in the eccentricity evolution as shown in Figure \ref{aerotM004_Mp1_eo001_ss}. The effect is even more significant for a BD dissipation of $1000~\sigma_{BD}$, indeed the eccentricity decreases quickly during the first $10^4$~yrs because both tides contribute to decreasing the eccentricity. At $t = 10^4$~yrs, $\Omega_{BD}/n$ becomes larger than $\frac{18}{11}\frac{Ne1(e)}{Ne2(e)}$, which defines the $\dot{e}=0$ line in a phase diagram. So the eccentricity increases and this effect is visible because the BD dissipation factor is high enough to make the BD tide dominant. At $t \sim 10^7$~yrs, the BD starts spinning up significantly, so the BD tide becomes negligible and the planetary tide contributes in making the eccentricity decrease once more.

%%%%%%%%%%%%%%%%%%%%%%%%%%%%%%%%

\subsubsection{Influence of the planetary dissipation}

We calculate the planetary dissipation using the estimate of \citet{DeSurgyLaskar1997} for the tidal lag time $k_{2,\oplus}\Delta t_{\oplus}$. Eqn (\ref{kdtsigma}) allows us to obtain the planetary dissipation factor as used by \citep{Hansen2010}. The planetary dissipation factor also depends strongly on the planetary radius: for a given mass and tidal dissipation factor, a larger planet experiences a stronger tidal flexion and dissipates more heat. 

The planetary tide is important when the eccentricity is non negligible. As the planet is in pseudo-synchronization it is constantly under changing stress which brings about more tidal dissipation. The system evolves to decrease the eccentricity to minimize the dissipation. 

	\begin{figure}[h!]
	\begin{center}
	\includegraphics[width=9cm]{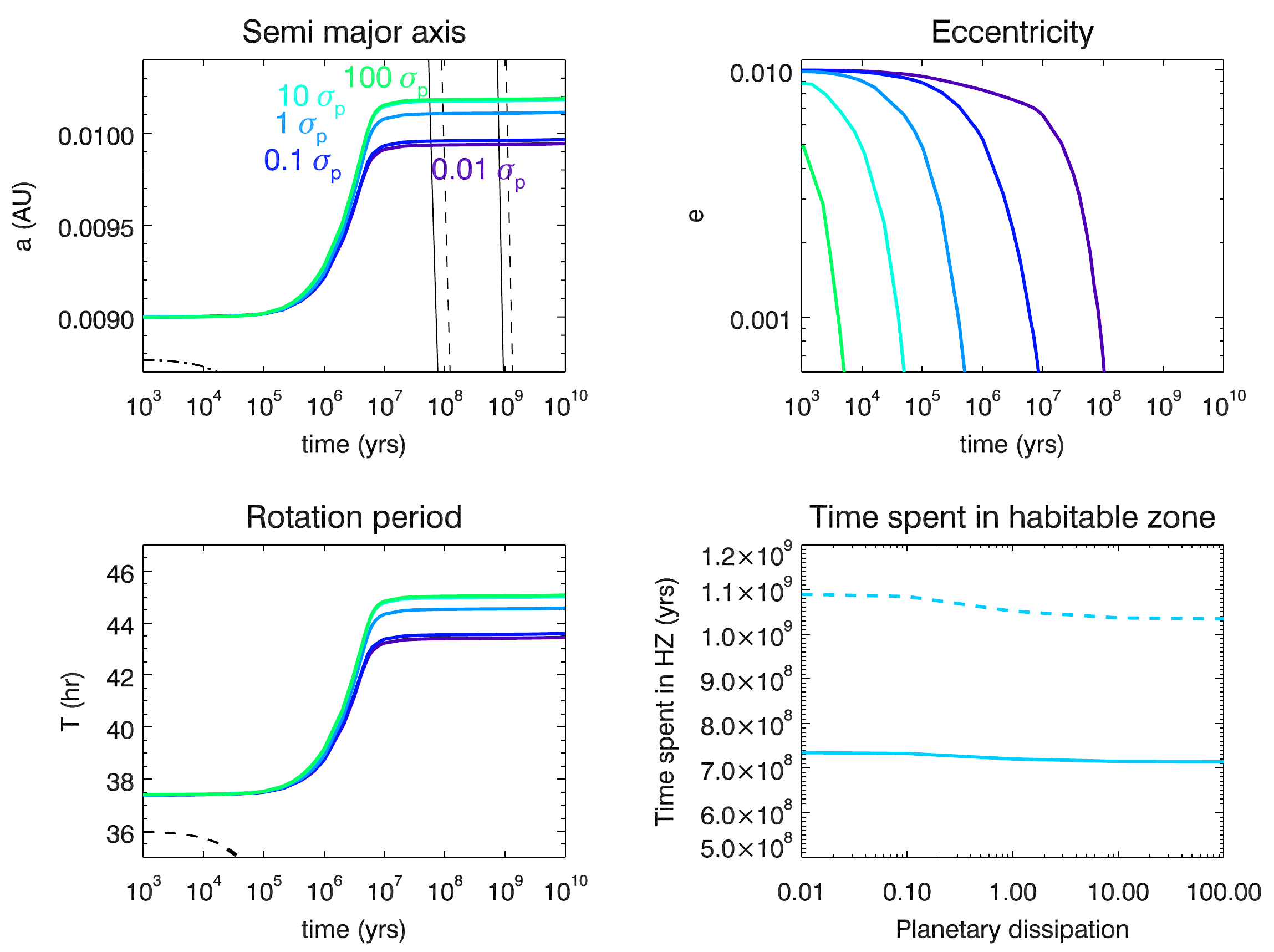}
	\caption{Tidal evolution of a $1~\Mearth$ planet starting at $9\times10^{-4}$~AU around a $0.04~\Msun$ BD for different planetary dissipation. Dependance on the planetary dissipation. The planet is pushed farther if the planetary dissipation is higher. The planetary dissipation factors are : $0.01$, $0.1$, $1$, $10$, $100\times\sigma_{p}$}
	\label{aerotM004_Mp1_eo001_sp}
	\end{center}
	\end{figure}

Figure \ref{aerotM004_Mp1_eo001_sp} shows the influence of the planetary dissipation on the tidal evolution of a $1~\Mearth$ planet orbiting a $0.04~\Msun$ BD.  The eccentricity decreases faster when the planetary dissipation factor is higher because the tides act more quickly to minimize the stress in the planet. Indeed, when planetary dissipation increases, the dissipation timescale $T_p$ decreases.  Within just $1000$ years significant differences can be seen between planets with different values of $\sigma_p$. From an initial eccentricity of $0.01$, the planet with the largest dissipation factor had its eccentricity cut in half in $1000$~yrs, while the planet with the lowest dissipation factor needs $\sim10^8$~yrs for this to occur. 

Planets with higher planetary dissipation rates are pushed farther from the BD. This is due to the fact that for low planetary dissipation, the eccentricity remains non negligible for a long time. Because of this eccentricity, the planetary tide counteracts a portion of the BD tide and thus the planet is not pushed as far. However, this is a relatively weak effect because changing the planetary dissipation by several orders of magnitude barely affects the final orbital distance. Indeed, for a planet beginning its evolution at an initial semi-major axis of $9\times10^{-3}$~AU, the final positions differ by only $2.5\%$ after $10$~Gyrs. If the planet initial semi-major axis is $12\times10^{-3}$~AU, they differ by only $0.5\%$. 

The planetary dissipation factor has little influence on the final semi-major axis for low initial eccentricity. The fact that the planet's orbit expands is almost entirely due to the BD tides. This has important implications for observations because if a planet is discovered orbiting a BD and its mass and orbit are known, then we can infer a value for the BD dissipation. However the difficulty in making that assessment is that it is only true for a system in which the eccentricity remained small throughout the evolution. Observing at time $t$ does not tell us about the evolution of the eccentricity. A combination of the initial eccentricity and the planetary dissipation can be very roughly estimated in systems with multiple planets (e.g., \citep{PapaloizouTerquem2010, Barnes2009a}) but that is beyond the scope of this paper.

	\begin{figure}[h!]
	\begin{center}
	\includegraphics[width=9cm]{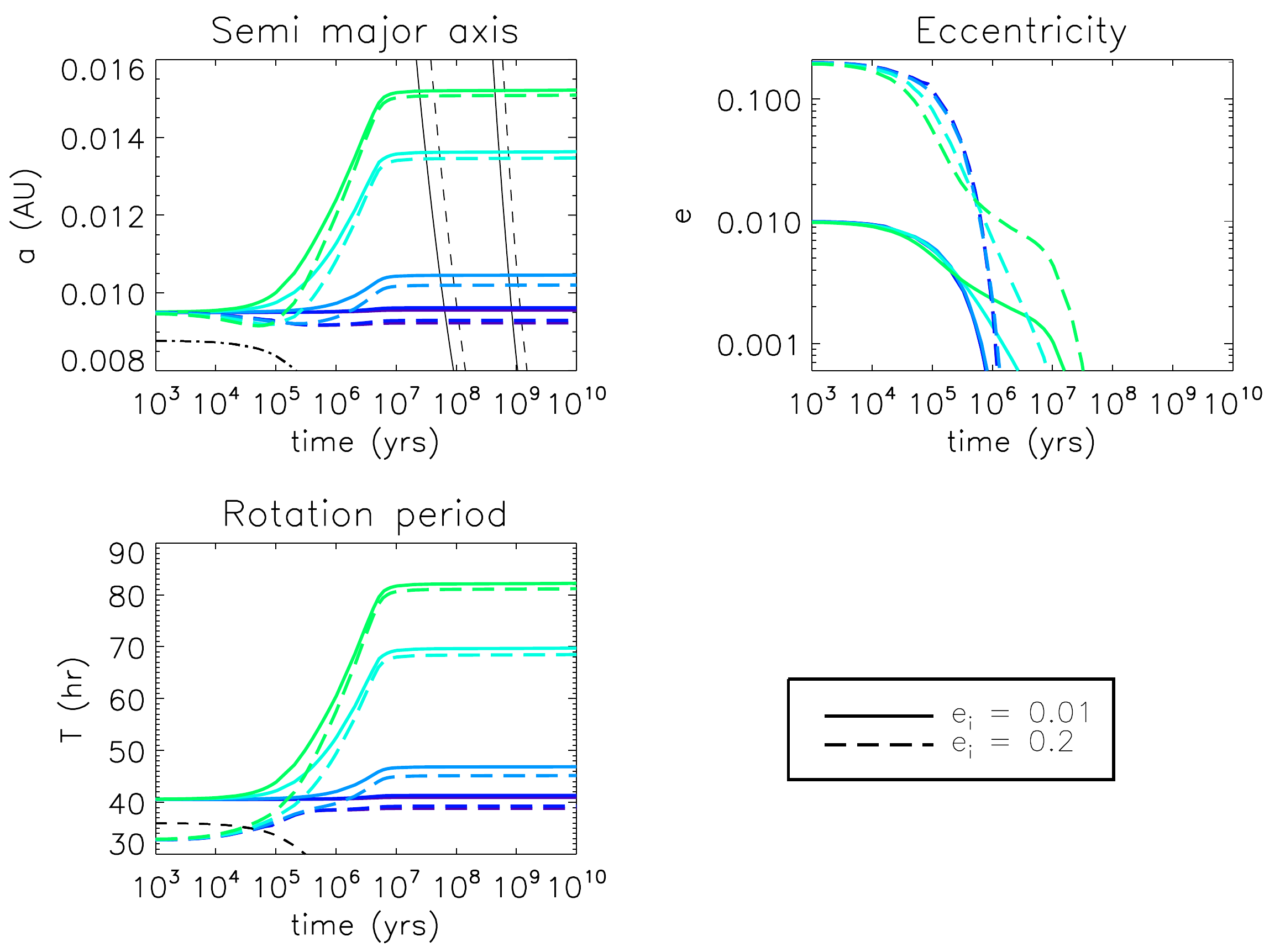}
	\caption{Tidal evolution of a $1~\Mearth$ planet starting at $9\times10^{-4}$~AU around a $0.04~\Msun$ BD for different BD dissipation and eccentricity. The BD dissipation factors are : $0.05$, $0.1$, $1$, $10$, $20\times\sigma_{\ast}$ and the initial eccentricity $0.01$ and $0.2$.}
	\label{aerotM004_Mp1_eo_ss}
	\end{center}
	\end{figure}	

	\begin{figure}[h!]
	\begin{center}
	\includegraphics[width=9cm]{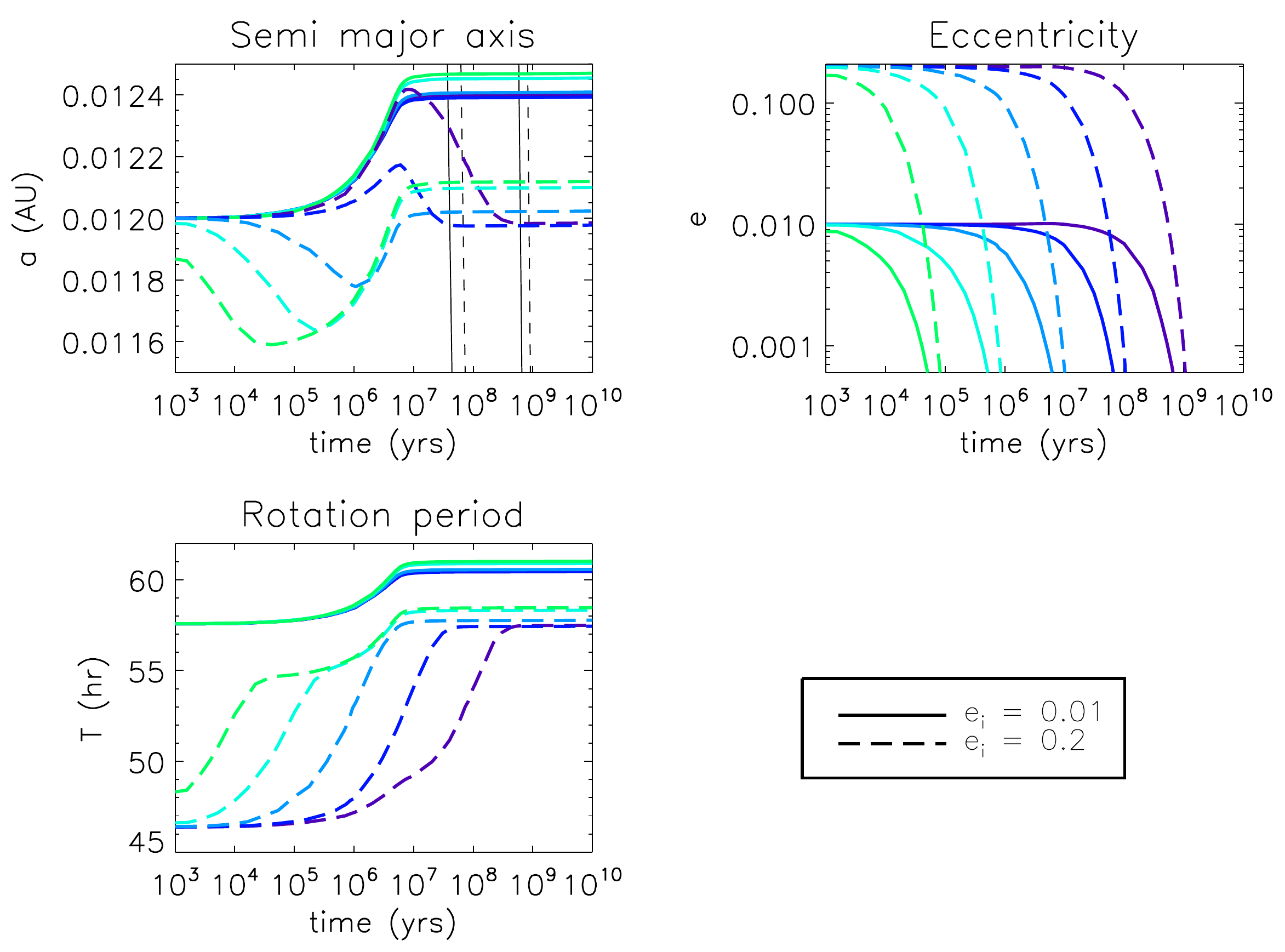}
	\caption{Tidal evolution of a $1~\Mearth$ planet starting at $9\times10^{-4}$~AU around a $0.04~\Msun$ BD for different planetary dissipation and eccentricity. The planetary dissipation factors are : $0.01$, $0.1$, $1$, $10$, $100\times\sigma_{p}$ and the initial eccentricity $0.01$ and $0.2$.}
	\label{aerotM004_Mp1_eo_sp}
	\end{center}
	\end{figure}

Figures \ref{aerotM004_Mp1_eo_sp} and \ref{aerotM004_Mp1_eo_ss} show the effect of varying both the initial eccentricity and the BD (Fig.  \ref{aerotM004_Mp1_eo_sp}) and planetary (Fig.  \ref{aerotM004_Mp1_eo_ss}) dissipation factors. The initial eccentricity has little influence on the dependance on BD dissipation because the eccentricity evolution is mainly determined by planetary tides. The eccentricity decreases to near zero in $10^6$~yrs to a few times $10^7$~yrs.  

However, varying the initial eccentricity has a strong influence on the effect on planetary dissipation. The eccentricity evolution itself is straightforward: $e$ decreases to negligible values in a few $10^4$~yrs for the highest planetary dissipation factors to $10^9$~yrs for the lowest dissipation factors. Higher eccentricity decreases faster than smaller ones so that for a same planetary dissipation factor both eccentricities reaches low values at about the same time. 

What is particularly interesting is that the initial eccentricity influences the final semi-major axis for different planetary dissipation rates.  For low planetary dissipation, the eccentricity remains non negligible for a long time and the evolution is complicated. The planetary tide is weak so the BD tide acts to slowly push the planet away, but when after a few million years the BD has shrunk significantly the BD tide's importance decreases to the point where the planetary tide eventually wins over and pushes the planet inwards.

If the planet is in pseudo-synchronization, the contribution of the planetary tide always pushes the planet inward and circularizes the orbit (see Fig. \ref{phasediag}). The only difference between low dissipation and high dissipation planets is then the timescale of the evolution. For higher planetary dissipation, the planetary tide is strong enough to quickly decrease the eccentricity and the evolution is then determined by the BD tide, whereas for lower planetary dissipation the eccentricity remains high for a long time and only at late times, when the BD tide becomes negligible, is the evolution determined by the planetary tide.

%%%%%%%%%%%%%%%%%%%%%%%%%%%%%%%%%%%%%%%%%%%%%%%%%%%%%%%%%%%%%%%
%%%%%%%%%%%%%%%%%%%%%%%%%%%%%%%%%%%%%%%%%%%%%%%%%%%%%%%%%%%%%%%

\section{Discussion}
\label{Discussion}
%%%%%%%%%%%%%%%%%%%%%%%%%%%%%%%%%%%%%%%%%%%%%%%%%%%%%%%%%%%%%%%

\subsection{Brown dwarfs and habitable zone}
\label{part3:1}

We now consider the habitability of planets orbiting BDs in the context of the circumstellar habitable zone, i.e. the range of orbital radii for which a planet's surface could have liquid water \citep{Kasting1993, Selsis2007}.  The issue of habitability is particularly interesting in the context of BDs because, as the BD contracts and cools the habitable zone moves inward.  Meanwhile, in most cases tidal evolution push planets outward.  Given this divergent radial evolution, planets that survive around BDs pass through the habitable zone and will thus spend a finite time with potentially habitable conditions.  \citet{Andreeshchev2004} found that for BD masses larger than $0.04~\Msun$ a planet can remain in the habitable zone for more than $4$~Gyrs if it forms sufficiently close to the Roche limit. However, \citet{Andreeshchev2004} did not take tidal evolution into account.  As we will see below, tidally-driven outward migration reduces this habitable time considerably.  

\medskip

We calculated the location of the habitable zone in two different ways. We first assumed that a planet on the outer edge receives $400$~W/m$^{2}$, the same flux as the $\sim 8 \Mearth$ potentially-habitable exoplanet GJ 581d \citep{Udry2007, Wordsworth2011}. The inner edge of the habitable zone  is taken to be four times this flux, $1600$~W/m$^{2}$; although this is not very physical it offers a reasonable approximation. The $400-1600$~W/m$^{2}$ habitable zone is marked with solid lines in Figs. \ref{aM004_Mp1_eo001}, \ref{aerotM004_eo001_Mp}, \ref{aerotM004_eo001_Msat}, \ref{aerotM004_eo001_Mjup}, \ref{aerotM004_Mp1_eo001_ss}, \ref{aerotM004_Mp1_eo001_sp}, \ref{aerotM004_Mp1_eo_ss} and \ref{aerotM004_Mp1_eo_sp}. Here the most important limit is the outer limit because it determines the time a planet spends on the HZ. For our second estimate we used the same method but assumed that a planet on the outer edge receives $240$~W/m$^{2}$, which is thought to be the minimum flux a planet requires to be habitable with the best atmospheric conditions possible. This second habitable zone estimate, of $240-480$~W/m$^{2}$ -- is shown with the dashed line in Figs. \ref{aM004_Mp1_eo001}, \ref{aerotM004_eo001_Mp}, \ref{aerotM004_eo001_Msat}, \ref{aerotM004_eo001_Mjup}, \ref{aerotM004_Mp1_eo001_ss}, \ref{aerotM004_Mp1_eo001_sp}, \ref{aerotM004_Mp1_eo_ss} and \ref{aerotM004_Mp1_eo_sp}.  We note that, given the diverse ways to heat a planet (e.g.,\citet{ForgetPierrehumbert1997}, \citet{PierrehumbertGaidos2011}), the outer edge of the habitable zone is in reality a function of at least planet's atmospheric properties.

We used the same tidal calculation presented in section 3 to study the influence of different parameters on how much time a planet spends in the habitable zone (see Figs. \ref{aerotM004_eo001_Mp}, \ref{aerotM004_Mp1_eo001_ss} and \ref{aerotM004_Mp1_eo001_sp}).  

Planets at larger orbital distances spend less time in the habitable zone because it sweeps past them early in their evolution.  Increasing the planet mass reduces the time spent in the habitable zone (Fig. \ref{aerotM004_eo001_Mp} ). Lighter planets are less influenced by tides so they will tend to have an almost constant semi-major axis. For a given initial semi-major axis, lighter planets spend more time in the habitable zone than heavier planets.  Moreover, if planets form with substantial eccentricity they receive more flux early in their evolution, which is not in favor of habitability. Planets in their early evolution are inside the inner edge of the HZ so they are already too hot for liquid water, a non zero eccentricity will just worsen the situation. For standard dissipation factors and for planets initially close to their BD, the eccentricity drops to near-zero before the planet reaches the habitable zone, or rather before the habitable zone reaches the planet. However, there do exist cases where the planet starts farther from the BD so the eccentricity remains high for a long time, which can contribute to keeping the planet habitable for a slightly longer time. Indeed, the flux received by the BD is $F(a,e\not=0) = F(a,e=0)(1-e^2)^{-1/2}$ \citep[see also][]{Barnes2009b}. However, the flux increase is only significant for very eccentric planets ($e \gtrsim 0.8$), and such planets would need to be relatively distant from their BD to maintain a high eccentricity high for a long time.

Increasing the BD dissipation and the planetary dissipation also decreases the time a planet spends in the habitable zone because the planets are pushed farther and the HZ sweeps past them early in their evolution. However, the effect of the BD dissipation is more pronounced than of the planetary dissipation, which is seen in Figs. \ref{aerotM004_Mp1_eo001_ss} and \ref{aerotM004_Mp1_eo001_sp}. The time spent in the habitable zone decreases faster if we increase the BD dissipation factor rather than the planetary dissipation factor. 

The main parameter that determines a planet's habitable zone lifetime is the mass of the BD. More massive BDs are more likely to host a habitable planet than low mass BDs. This is due to the evolution of the luminosity as a function of BD mass. The luminosity of high mass BDs is higher and decreases less than for low mass BDs. So the HZ for high mass BDs is farther from the BD and this is favorable for habitability because tidal evolution tends to be weaker such that outward migration is slower and less pronounced.

	\begin{figure}[h!]
	\begin{center}
	\includegraphics[width=8cm]{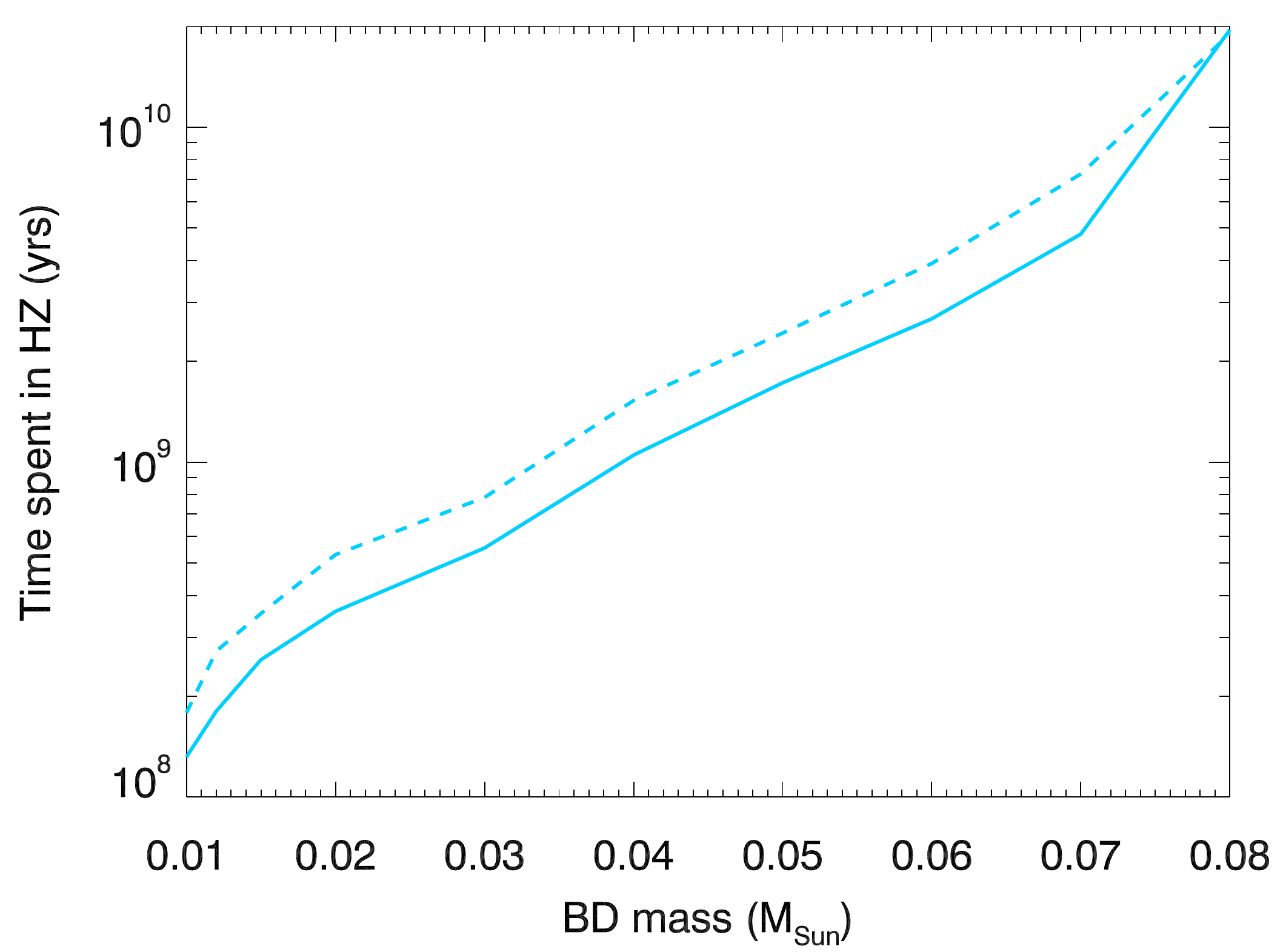}
	\caption{Time spent in the HZ by a $1~\Mearth$ orbiting a $0.04~\Msun$ BD. Time zero is $1$~Myr. This time has been calculated with the most optimistic case, which is for $a_{switch}$.}
	\label{TimeHZMs_106}
	\end{center}
	\end{figure}

To sum up, the most optimistic scenario for habitability exists for BDs with masses above $0.04~\Msun$ and for parameters that induce the smallest orbital change, that is low BD and planetary dissipation and low planetary mass. However, it is important to keep in mind that the final position of planets and thus the time they spend in the HZ depends predominantly on the initial semi-major axis compared to the corotation radius. Figure \ref{rot_all106_eo001_ss} shows that for low BD dissipation, the best case corresponding to $a_{switch}$ leads to orbital changes and actually brings the planet closer to the BD than it was initially.

%%%%%%%%%%%%%%%%%%%%%%%%%%%%%%%%%%%%%%%%%%%%%%%%%%%%%%%%%%%%%%%

\subsection{Observational possibilities}

BDs are thought to be numerous and their small masses and radius makes them better targets for radial velocity and transit searches than main sequence stars for a given limiting magnitude \citep{Blake2008}. 

Observing planets orbiting BDs could help constrain tidal dissipation models. In Section 3 we showed that the final orbital distance of a planet depends mainly on its initial position compared to the initial corotation radius. It is therefore hard to predict anything from a single observation of a planet orbiting a BD. However, a large sample of planets orbiting BDs could place constraints on certain tidal parameters.

	\begin{figure}[h!]
	\begin{center}
	\includegraphics[width=9cm]{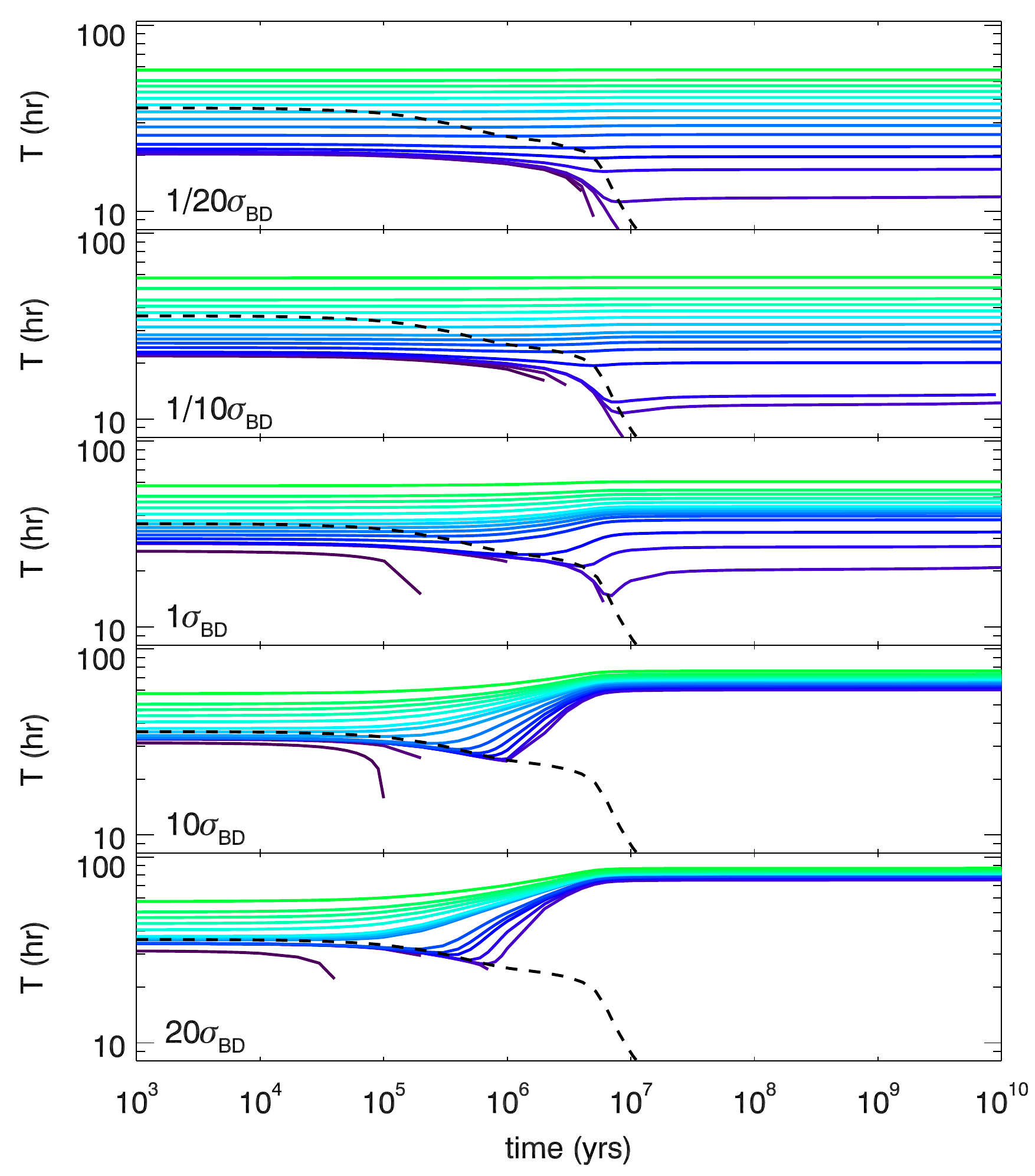}
	\caption{{\bf Tidal} evolution of the rotation period a $1~\Mearth$ planet orbiting a $0.04~\Msun$ BD. The different panels are for different BD dissipation factors : $0.05$, $0.1$, $1$, $10$, $20\times\sigma_{BD}$. The initial eccentricity is $0.01$.}
	\label{rot_all106_eo001_ss}
	\end{center}
	\end{figure}

The main quantity that could be constrained with a sample of planets orbiting BDs is the $\sigma_{BD}$, the dissipation rate within the BD. Figure \ref{rot_all106_eo001_ss} shows the evolution of the rotation period of a suite of $1~\Mearth$ planets orbiting a $0.04~\Msun$ BD. The eccentricity drops to zero and the planet is in true spin-orbit synchronization, so the y-axis in Fig. \ref{rot_all106_eo001_ss} represents the orbital period. As we saw in subsection \ref{Infstdiss}, the BD dissipation factor strongly influences the final planetary orbit.

Fig. \ref{rot_all106_eo001_ss} shows that for low BD dissipation the distribution of final orbital distances is less densely packed than the distribution of initial orbital distances, with orbital periods as small as $\sim 10$~hrs. The planets' evolution is dominated by planetary tides, which bring the planets closer to the BD and effectively spread the initial planet distribution.  In contrast, for the highest BD dissipation the distribution of final orbital distances is much more densely packed than the distribution of initial orbital distances and at much larger orbital periods because the BD tide dominated and planets have all been pushed outward. The planets accumulate at an orbital period of $\sim 80$~hrs. 

	\begin{figure}[h!]
	\begin{center}
	\includegraphics[width=9cm]{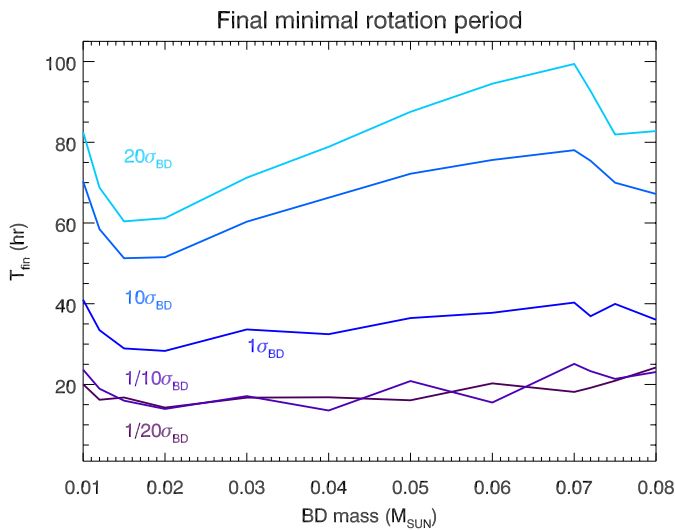}
	\caption{Final minimal period of $1~\Mearth$ planets as a function of the BD mass and for different BD dissipations.}
	\label{rot_finall_Ms_ss}
	\end{center}
	\end{figure}

Figure \ref{rot_finall_Ms_ss} shows the minimum final orbital period for $1~\Mearth$ planets as a function of the BD mass and the BD dissipation.  As expected, the minimum period increases for higher BD dissipation. For low BD dissipation, the minimal period is about $20$~hrs and does not change much with mass. The fluctuations observed are due to the extreme sensitivity to initial conditions because the final orbital distance of the planets depends on when they cross the shrinking corotation distance. However, the decrease of the final rotation period from a BD mass of $0.01~\Msun$ to $0.02~\Msun$ was explained in subsection \ref{massstar} and is due to the fact a BD of $0.01~\Msun$ pushes the planets during its entire evolution. For dissipations of $1$, $10$ and $20~\sigma_{BD}$, the minimal final period increases between masses of $0.02~\Msun$ and $0.07~\Msun$ because the more massive the BD, the stronger the BD tide. However the final decrease between $0.07~\Msun$ and $0.08~\Msun$ is due to sensitivity to initial conditions. For example, for $\Ms = 0.07~\Msun$, the planet crosses the corotation radius at a few $10^5$~yrs and for $\Ms = 0.072~\Msun$ it crosses it at $\sim 10^6$~yrs. So the planet around the $0.07~\Msun$  BD is pushed outward sooner when the BD radius is still large so the planet is pushed more strongly and for a longer time.  The precision to which $a_{switch}$ has been determined here is not sufficient to remove these fluctuations . 

Thus, with a sufficiently large sample of planets orbiting BDs with well-characterized masses and orbits, it should be possible to infer -- or at least constrain -- values for the dissipation in BDs.

Figures \ref{rot_all106_eo001_ss} and \ref{rot_finall_Ms_ss} assumed a time zero of $1$~Myr.  However, as seen in Section \ref{inftime}, tidal evolution is weaker if time zero is later simply because BDs contract significantly during $1$ and $10$~Myr.  Planets can form and survive closer to their BD hosts if time zero is later, i.e. for longer lifetimes of the gaseous protoplanetary disk.  Thus, planets may have even shorter orbital periods than those shown in Fig. \ref{rot_finall_Ms_ss}â for later time zero. For a $1~\Mearth$ planet orbiting a $0.04~\Msun$ BD and a time zero of $10$~Myr, the final minimal periods are of about $10$~hr for BD dissipation of $0.05$ and $0.1~\sigma_{BD}$, $20$~hr for $1~\sigma_{BD}$, $30$~hr for $10~\sigma_{BD}$ and $40$~hr for $20~\sigma_{BD}$. If time zero is $10$~Myr, the final minimal period is half it is if time zero is $1$~Myr.

%%%%%%%%%%%%%%%%%%%%%%%%%%%%%%%%%%%%%%%%%%%%%%%%%%%%%%%%%%%%%%%
%%%%%%%%%%%%%%%%%%%%%%%%%%%%%%%%%%%%%%%%%%%%%%%%%%%%%%%%%%%%%%%

\section{Conclusion}

In this paper, we calculated the tidal evolution of planets around BDs using a standard constant time lag tidal model (e.g., Leconte et al 2010, Hansen 2010).  Our calculations account for the fact that BDs contract with time \citep{Baraffe2003} and therefore spin up. We calibrated the BD initial rotation period with observational data \citep{Herbst2007}. We then investigated the parameter space in order to gauge each parameter's effect on the evolution of planets orbiting BDs and on the time planets can spend in the habitable zone. 

We found that the most important parameter was the initial orbital distance compared to the corotation distance. A planet outside the corotation radius migrates outward and stops at a distance where tidal effects are negligible. A planet inside the corotation radius evolves inward and is likely to eventually merge with the BD.  However, as the BD spins up with time the corotation radius shrinks and in some cases the corotation radius can pass by the planet orbit and reverse the direction of the migration. The fact that BDs contract with time allows the system to experience many different evolutions given the importance of parameters such as the initial eccentricity and the dissipation factors of the planet and the BD. The planetary tide and the BD tide vary in intensity and the value of their ratio is extremely helpful to understand which behavior is due to which tide. 

Thus the study of tidal evolution of a planet orbiting a BD is interesting from a theoretical point of view because of the variety of possible evolutions due to the spin-up of the BD and the effect of parameters like the dissipation factors. But it is also interesting from an observational point of view because planets that form close to their BD can have very short final orbital periods. In the most favorable cases, their periods can be as short as $\sim 10$~hrs. Those planets could be observed in transit around BDs. 

For BDs more massive than 0.05~M$_\odot$, planets can spend  a significant amount of time in the habitable zone - from $1$~Gyr to $\sim 10$~Gyrs- so this raises also the question of habitability of such planets. Furthermore, planets with pseudo-synchronous or synchronous rotation are interesting case studies for climate modeling \citep{Wordsworth2011}.

%%%%%%%%%%%%%%%%%%%%%%%%%%%%%%%%%%%%%%%%%%%%%%%%%%%%%%%%%%%%%%%

\begin{acknowledgements}
We thank Franck Selsis, Adrian Belu and Amaury Triaud for their ideas and support.  We also thank the CNRS's PNP program for funding, and the Conseil Regional d'Aquitaine for help purchasing the computers on which these calculations were performed.  We also acknowledge funding from the European Community via the P7/2007-2013 Grant Agreement no. 247060. We thank the referee Brad Hansen for the useful comments.
 \end{acknowledgements}

%%%%%%%%%%%%%%%%%%%%%%%%%%%%%%%%%%%%%%%%%%%%%%%%%%%%%%%%%%%%%%%

%\bibliographystyle{aa}

%\bibliography{aamnem99,biblio}

\end{document}